\documentclass{aa}
\begin{document} 

   \title{Solar wind dynamics around a comet}
   \subtitle{The paradigmatic inverse-square-law model}
   \titlerunning{The paradigmatic inverse-square-law model}
   
   \author{M. Saillenfest\inst{1}, B. Tabone\inst{2}, E. Behar\inst{3,4}}
   \institute{IMCCE, Observatoire de Paris, PSL Research University, CNRS, Sorbonne Universit\'e, UPMC Univ. Paris 06, LAL, Université de Lille, 75014 Paris, France
              \and
              LERMA, Observatoire de Paris, PSL Research University, CNRS, Sorbonne Universit\'e, UPMC Univ. Paris 06, 75014 Paris, France
              \and
              Swedish Institute of Space Physics, Kiruna, Sweden
              \and
              Lule\aa\ University of Technology, Department of Computer Science, Electrical and Space Engineering, Kiruna, Sweden\\
              \email{melaine.saillenfest@obspm.fr}
              }

   \date{Received 01 February 2018; Accepted 22 May 2018}
 
   \abstract
   {}
   {Observations of solar protons near comet 67P/Churyumov-Gerasimenko (67P) by the \emph{Rosetta} spacecraft can be modelled by the planar motion in an effective magnetic field proportional to $1/r^2$. We aim to provide a thorough study of such dynamics, with a clear description of the behaviour of an incoming flux of particles. We will be able, then, to calibrate the free parameters of the model to \emph{Rosetta} observations.}
   {Basic tools of dynamical analysis are used. They lead to a definition of the relevant parameters for the system and a classification of the possible types of trajectories. Using the so-obtained formalism, the structures formed by a flux of particles coming from infinity can be studied.}
   {All the trajectories are parametrised by two characteristic radii, $r_E$ and $r_C$, derived from first integrals. There are three different types of motion possible divided by a separatrix corresponding to $r_E=r_C$. An analytical expression of the trajectories, defined by an integral, is developed. Using this formalism, the application to a flux of particles coming from infinity (modelling the incident solar wind) gives one free parameter only, the radius $r_E$, which scales the problem. A circular cavity of radius $0.28\,r_E$ is created, as well as an overdensity curve (analogous to a caustic in optics). At each observation time, $r_E$ can be calibrated to \emph{Rosetta} plasma measurements, giving a qualitative understanding of the solar particle dynamics (incoming direction, cavity and density map). We also deduce that, in order to properly capture the essence of the dynamics, numerical simulations of the solar wind around a comet must use simulation boxes much larger than $r_E$ and grids much finer than $r_E$.}
   {}
   
   \keywords{solar wind -- coma -- magnetic field}

   \maketitle

\section{Introduction}
   Plasma instruments on board the \emph{Rosetta} mission have provided invaluable information about the dynamics of solar and cometary ions in a comet neighbourhood (comet 67P/Churyumov-Gerasimenko, or 67P). For the first time, these dynamics were followed as the comet nucleus activity was evolving, from more than $3.8$~astronomical units (au) to its perihelion at $1.2$~au. Because of the lower production rate of 67P's nucleus together with these large heliocentric distances, the interaction between the solar wind and the cometary atmosphere (coma), was fundamentally different from what was previously observed at more active comets closer to the Sun \citep[comet Halley or comet Giacobini-Zinner:][]{GREWING-etal_1988,COWLEY_1987}. Because the gyration scale of the cometary ions is larger than the interaction region, the classical fluid treatment of the plasmas does not apply at 67P, and a kinetic description of the interaction is necessary. In \citet{BEHAR-etal_2018a} the terms ``fluid comet'' and ``kinetic comet'' were introduced to separate these two different regimes. As of now, only self-consistent numerical models have tackled the interaction between the solar wind and the coma of a ``kinetic'' comet \citep{BAGDONAT-MOTSCHMANN_2002, HANSEN-etal_2007, RUBIN-etal_2014b, KOENDERS-etal_2016a, KOENDERS-etal_2016b, BEHAR-etal_2016, DECA-etal_2017, HUANG-etal_2018}. All these models result in a highly asymmetric plasma environment, in contrast with the classical symmetric picture obtained for more active comets closer to the Sun \citep{RUBIN-etal_2014a}.

   In the context of comet 67P and based on in-situ data, \cite{BEHAR-etal_2017} recently showed that a cavity completely free of solar particles is created around the comet's nucleus, surrounded by a region where they are focused in a specific direction. Moreover, the measured velocity of the solar protons is almost constant in norm throughout the mission, indicating that they are deflected without significant loss of energy. Remarkably, the main plasma observations are very well reproduced by a simple inverse-square-law ``effective magnetic field'' applied to the incoming flux of solar protons. Using this model, the density and velocity profiles become natural geometrical effects, also in qualitative agreement with numerical simulations \citep{BEHAR-etal_2016,BEHAR-etal_2017}. Due to the striking success of this empirical approach, it became necessary to outline the intrinsic properties of this force field: this will allow us to state clearly what it would imply for the dynamics of solar wind protons, and hopefully to link the observables to physical quantities.
   
   The aim of this paper is to provide a full characterisation of the planar dynamics of charged particles in a magnetic field proportional to $1/r^2$. This way, the appropriate formalism will be available for further applications to the \emph{Rosetta} mission or any analogous physical modelling. In particular, the behaviour of an incoming flux of charged particles in a $1/r^2$ magnetic field has only been explored by numerical means so far, and its precise characteristics are still missing. Consequently, this paper is mainly devoted to dynamical aspects, and we will only hint at the physical considerations regarding its application to comets. Crucial discussions about the nature of this force and comparisons to self-consistent physical models of comet 67P are presented in companion papers \citep{BEHAR-etal_2018a,BEHAR-etal_2018b}.

   The planar dynamics in an inverse-cube-law magnetic field has been thoroughly studied by physicists because it describes the motion of charged particles in the equatorial plane of a magnetic dipole \citep[e.g.][]{STORMER_1907,STORMER_1930,GRAEF-KUSAKA_1938,LIFSHITZ_1942,DE-VOGELAERE_1950,AVRETT_1962}. The interest for such dynamics was greatly enhanced by its direct applications to the geomagnetic field. Unbounded and bounded solutions exist and trapped particles are indeed observed around the Earth \citep{WILLIAMS_1971}. The deflection of an incoming flux of particles seems to produce similar structures as for an inverse-square law (compare Fig.~1 by \citealp{SHAIKHISLAMOV_2015} with Fig.~3 by \citealp{BEHAR-etal_2017}). Besides, complex plasma interactions in other physical contexts could possibly be modelled as well by such simple laws. A comparative study of the different powers of $1/r$ would thus be also valuable.

   This paper is organised as follows: Section~\ref{sec:dyn} presents a general study of the inverse-square-law magnetic field. After having summarised the model developed by~\cite{BEHAR-etal_2018a}, we define the different types of possible orbits, outline their properties, and give for them an analytical expression defined by an integral. In Sect.~\ref{sec:flux}, we apply this formalism to an incoming flux of particles, similar to the solar protons. The properties of the cavity and of the overdensity region reported by \cite{BEHAR-etal_2017} are fully characterised. Then, Sect.~\ref{sec:rosetta} presents order-of-magnitude estimates of the characteristic quantities of the model calibrated on the plasma observations realised by the \emph{Rosetta} spacecraft.
   
   Additionally, the comparison of dynamics produced by magnetic fields proportional to an arbitrary power of $1/r$ is given in Appendix~\ref{sec:comp}: it could serve as reference when dealing with analogous problems.
   
\section{General study of the dynamics}\label{sec:dyn}
   
   \subsection{The inverse-square law for solar protons around comets}\label{ssec:model}
   The analytical model introduced by \cite{BEHAR-etal_2018a} is built from three sub-models. We outline here their main characteristics (readers mainly interested in dynamical aspects can safely go to Sect.~\ref{ssec:eqmot}).
   
   Steady state is always assumed, implying that the change of heliocentric distance of the comet is slow enough to be considered as an adiabatic process. The first sub-model is a description of the ionised coma and its density distribution. The cometary atmosphere is assumed to have a spherical symmetry: the neutral elements are produced at a rate $Q$ and expand radially in all directions with constant velocity $u_0$. The cometary ions are created from these neutral elements with a rate $\nu_i$ (mainly by photo-ionisation and electron-impact ionisation). They initially have the radial velocity $u_0$, but they are accelerated by the local electric and magnetic fields and lost from the system. This is taken into account by a ``destruction'' rate $\nu_{ml}$ (where $ml$ stands for mass loading). Therefore, in the regime of the system under study, the local density of cometary ions can be written as
   \begin{equation}\label{eq:ncom}
      n_{com} = \frac{\nu_i}{\nu_{ml}}\frac{Q}{4\pi u_0R^2}\,,
   \end{equation}
   where $R$ is the radial distance from the nucleus \citep[see][for details]{BEHAR-etal_2018a}. In this description, the ionised component of the coma is essentially made of the slow, new-born cometary ions, which are steadily created and lost. The second sub-model is a description of the magnetic field piling up due to the local decrease in the average velocity of the electrons (as slow new-born ions are added to the flow). The magnetic field $\mathbf{B}$ is considered frozen in the electron fluid, the latter coming from infinity on parallel trajectories. The third sub-model is a description of the electric field, which is reduced to its main component, the so-called motional electric field,
   \begin{equation}
      \mathbf{E} = -\underline{\mathbf{u}_i}\times\mathbf{B}\,,
   \end{equation}
   where $\underline{\mathbf{u}_i}$ is the average velocity of all charges carried by solar and cometary ions. Considering only the Lorentz force, this results in a generalised gyromotion for both populations, where the two gyroradii depend strongly on the density ratio. This generalised gyromotion is the core of the model, giving a mechanism through which energy and momentum are transferred from one population to the other. For simplicity, we finally consider that the cometary particles are mainly composed of water, resulting in the same charge $+e$ for the cometary ($H_2O^+$) and solar wind ($H^+$) ions. Putting all things together, the force applied to the solar protons is
   \begin{equation}
      m\,\ddot{\mathbf{x}} = e\,\frac{n_{com}}{n_{sw}}\,\dot{\mathbf{x}}\times\mathbf{B}_\infty\,.
   \end{equation}
   In this expression, $\mathbf{x}$ is the position vector of the proton, $m$ is its mass, and the dot means time derivative. The magnetic field $\mathbf{B}_\infty$ is the one carried by the solar wind before its encounter with the comet, and $n_{sw}$ is the average density of solar wind protons. Injecting the expression of $n_{com}$ from Eq.~\eqref{eq:ncom}, we finally get an inverse-square law like the one studied in the rest of this article.
   
   It should be noted that the force applied here to solar wind protons is not a magnetic field as such, but it behaves like one. Hence, even if we speak generically of ``magnetic field'' throughout this article, the reader should understand ``effective magnetic field'' to mean a vector field behaving as a magnetic field but possibly produced as a result of more complex interactions. In our case, the relevant dynamics takes place in a plane, written $(x,y)$ in the following. In the comet-Sun-electric frame (CSE) used by \cite{BEHAR-etal_2017}, this plane contains the comet, the Sun, and the electric field vector produced by the incoming solar wind.\footnote{The axes in the plane of motion are labelled $(x,z)$ in \cite{BEHAR-etal_2017,BEHAR-etal_2018a}, which is the traditional convention used in solar wind studies (the effective magnetic field is thus oriented along the $y$ axis). We think that the notation $(x,y)$ is more appropriate for the present paper, focussed on dynamics only. This should not be too confusing for the reader.} During its operating phase around comet 67P, the \emph{Rosetta} spacecraft was not far from this plane (since it was not far from the comet itself). In other contexts, the $(x,y)$ plane used here could be the equatorial plane of some source of magnetic field.
   
   \subsection{Equations of motion}\label{ssec:eqmot}
   A particle of mass $m$, charge $q$, and position $\mathbf{x} = (x,y,z)^T$ is subject to a magnetic field of the form
   \begin{equation}\label{eq:B}
      \mathbf{B}(\mathbf{x})=\frac{\alpha}{x^2+y^2}
      \begin{pmatrix}
         0\\
         0\\
         1
      \end{pmatrix}
      ,\hspace{1cm}\text{where }\alpha\in\mathbb{R}.
   \end{equation}
   From the classical Lorentz force, the equations of motion are $m\,\ddot{\mathbf{x}} = q\,\dot{\mathbf{x}}\times\mathbf{B}(\mathbf{x})$, that is,
   \begin{equation}\label{eq:ini}
      \begin{pmatrix}
         \ddot{x} \\
         \ddot{y} \\
         \ddot{z}
      \end{pmatrix}
      = \frac{k}{x^2+y^2}
      \begin{pmatrix}
          \dot{y} \\
         -\dot{x} \\
         0
      \end{pmatrix}
      ,\hspace{1cm}\text{where }k=\frac{\alpha\,q}{m}\in\mathbb{R},
   \end{equation}
   in which the constant $k$ has the dimension of length time velocity and the dot means derivative with respect to the time $t$. From Eq.~\eqref{eq:ini}, the vertical velocity is constant and imposed by the initial conditions. We are interested here in the dynamics in the $(x,y)$ plane. Let us introduce the polar coordinates $(r,\theta)$. The equations of motion rewrite\footnote{We get here the same equations as~\cite{GRAEF-KUSAKA_1938}. This comes from a mistake in their paper: they begin with the equations of a $1/r^2$ field; they introduce the conserved quantities of a $1/r^3$ field; they write down equations mixing both types of fields, and eventually, they study the $1/r^3$ one for the rest of the paper. Since they deal with the motion in the equatorial plane of a magnetic dipole, $1/r^3$ is the correct field to use.} as
   \begin{align}
      \ddot{r}-r\dot{\theta}^2 &= \frac{k}{r}\dot{\theta} \label{eq:dyn1} \\
      r\ddot{\theta}+2\dot{r}\dot{\theta} &= -\frac{k}{r^2}\dot{r}\,. \label{eq:dyn2}
   \end{align}
   Since the force is always perpendicular to the velocity vector, its norm is constant (conservation of the total energy $E$). This leads to the first integral
   \begin{equation}
      v=\sqrt{\dot{r}^2+r^2\dot{\theta}^2}\,,
   \end{equation}
   equal to the norm of the velocity projected in the $(x,y)$ plane. Moreover, Eq.~\eqref{eq:dyn2} is directly integrable:
   \begin{equation}
      \frac{\mathrm{d}(r^2\dot{\theta})}{\mathrm{d}t} = -k\frac{\mathrm{d}\ln(r/r_\star)}{\mathrm{d}t},
   \end{equation}
   where the arbitrary constant $r_\star$ is added for dimensionality reasons. This leads to a second first integral,
   \begin{equation}
       c = r^2\dot{\theta} + k\ln(r/r_\star) = const.
   \end{equation}
   It can be thought as the conservation of a generalised angular momentum, coming from the symmetry of rotation around the $z$-axis.
   
   From dimensionality arguments, the conservation of $v$ makes a characteristic length and a characteristic frequency of the system appear:
   \begin{equation}\label{eq:norm}
      r_E = \frac{|k|}{v}
      \hspace{1cm};\hspace{1cm}
      \omega_E = -\frac{k}{r_E^2} = -\frac{v^2}{k}\,.
   \end{equation}
   In the same way, the generalised angular momentum $c$ can be turned into the characteristic length
   \begin{equation}
      r_C = r_\star\exp(c/k+1) = r\exp(r^2\dot{\theta}/k+1)\,.
   \end{equation}
   As we will see, the dynamics of the particle is entirely contained inside the independent constants $r_C$ and $r_E$. Their physical meaning will appear later.
   
   Similarly to \cite{STORMER_1930}, it is convenient at this point to use the normalised quantities
   \begin{equation}
      \rho=r/r_E
      \hspace{1cm};\hspace{1cm}
      \mathrm{d}\tau=\omega_E\,\mathrm{d}t.
   \end{equation}
   We note that if $k>0$, the direction of time is reversed. In the new coordinates, the equations of motion~(\ref{eq:dyn1}-\ref{eq:dyn2}) become
   \begin{align}
      \rho\ddot{\rho}-\rho^2\,\dot{\theta}^2 = -\dot{\theta} \label{eq:motion1}\\
      \rho^2\ddot{\theta}+2\rho\dot{\rho}\dot{\theta} = \frac{\dot{\rho}}{\rho}\,, \label{eq:motion2}
   \end{align}
   where this time the dot means derivative with respect to the normalised time $\tau$ (this double use of the dot should not be confusing for the reader, since $t$ is only used with the dimensional $r$ coordinate, while $\tau$ is only used with the dimensionless $\rho$ coordinate). Equations~(\ref{eq:motion1}-\ref{eq:motion2}) are equivalent to:
   \begin{align}
      \dot{\rho}^2+\rho^2\dot{\theta}^2 = 1 \label{eq:energy} \\
      \rho_C = \rho\exp(-\rho^2\dot{\theta}+1) \,, \label{eq:rho0}
   \end{align}
   coming respectively from the energy and the generalised angular momentum (we have in particular $\rho_C=r_C/r_E$).
   
   \subsection{Geometry of the trajectories}
   
   \begin{figure*}
      \centering
      \includegraphics[width=0.8\textwidth]{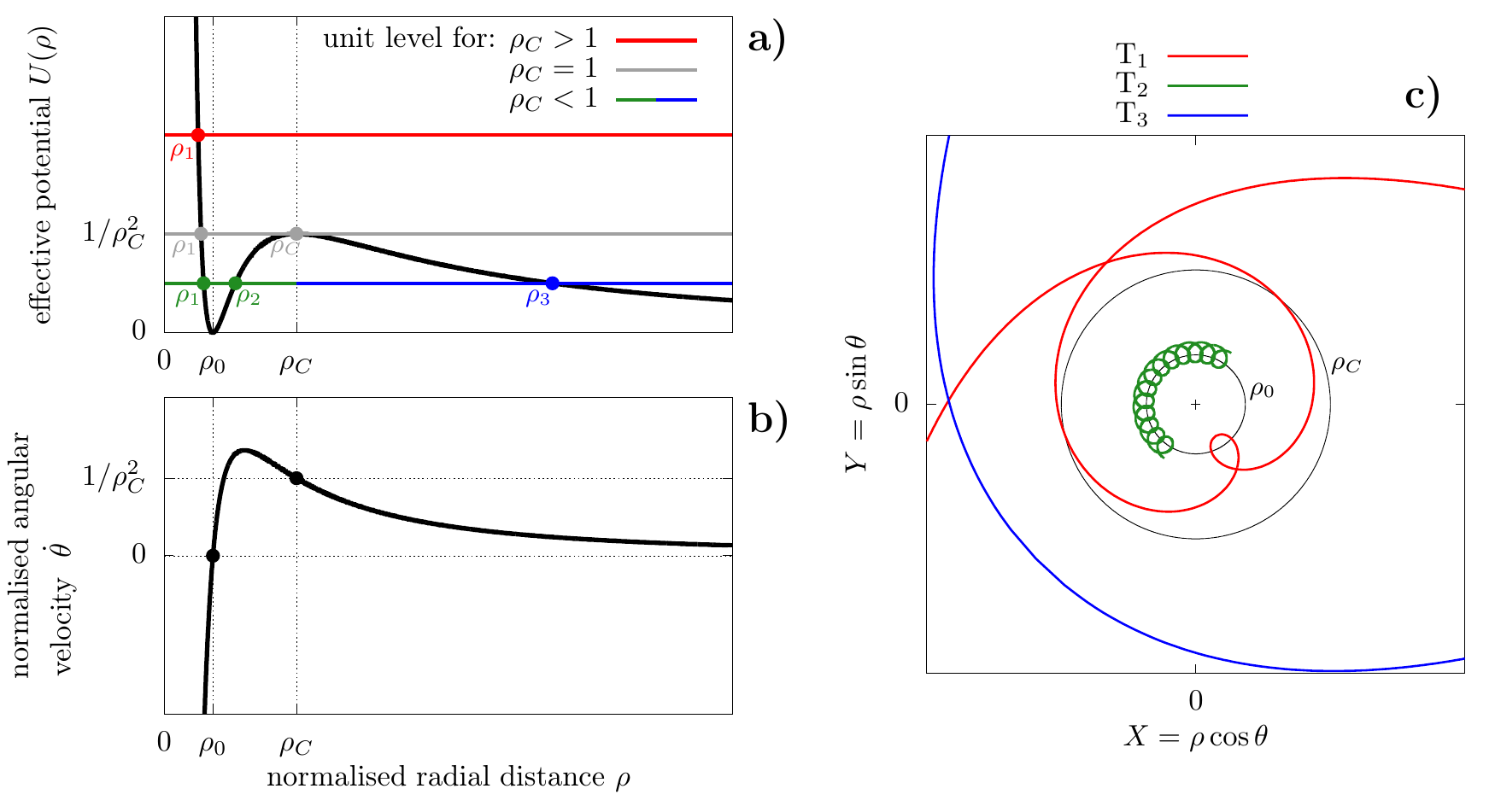}
      \caption{\textbf{a)} Effective potential as a function of $\rho$. Changing the value of parameter $\rho_C$ is equivalent to rescaling the axes. The unit level on the vertical axis gives the intervals of $\rho$ allowed for the particle, such that $U(\rho)<1$. These intervals are delimited by $\rho_1$, $\rho_2$, and $\rho_3$, given at Eq.~\eqref{eq:r1r2r3}. \textbf{b)} Angular velocity as a function of $\rho$. \textbf{c)} Examples of trajectories for three orbit types, obtained by using the expression from Eq.~\eqref{eq:traj} with parameters $\rho_C=(1.01,0.5,0.9)$ for (red, green, blue). The axes are rescaled such that $\rho_C$ appears the same, as in graph \textbf{a}.}
      \label{fig:poteff}
   \end{figure*}
   
   Introducing $\rho_C$ (Eq.~\ref{eq:rho0}) into the energy expression (Eq.~\ref{eq:energy}), we get
   \begin{equation}\label{eq:poteff}
      \dot{\rho}^2+U(\rho)=1
      \hspace{0.5cm}\text{with}\hspace{0.5cm}
      U(\rho) = \left(\frac{\ln(\rho/\rho_0)}{\rho}\right)^2\,,
   \end{equation}
   where we define
   \begin{equation}
      \rho_0 \equiv \rho_C\exp(-1)\,.
   \end{equation}
   We will see below that both $\rho_0$ and $\rho_C$ have a precise dynamical meaning. Since they are directly proportional, the problem can be indifferently parametrised by one or the other. For simplicity, we will consider either $\rho_0$ or $\rho_C$ in the following, according to the dynamical feature under discussion.
   
   The function $U$ can be interpreted as an effective potential, which counterbalances the kinetic term at all times. Its general form gives directly the values of $\rho$ allowed as a function of the parameter $\rho_C$ (Fig.~\ref{fig:poteff}a). Noting \{T$_i$\} the different types of trajectories, we have\footnote{In dimensional coordinates, the three cases correspond to $r_C>r_E$, $r_C<r_E$, and $r_C=r_E$.}
   \begin{equation}\label{eq:types}
      \begin{aligned}
         &\rho_C > 1\\
         &\hspace{0.4cm}
         \text{T$_1$ : unbounded orbit } (\rho\geqslant \rho_1)\\
         &\rho_C < 1\\
         &\hspace{0.4cm}
         \left\{
         \begin{aligned}
            &\text{T$_2$ : bounded orbit }(\rho_1\leqslant \rho\leqslant \rho_2) \\
            &\text{T$_3$ : unbounded orbit }(\rho\geqslant \rho_3)
         \end{aligned}
         \right.
         &\\
         &\rho_C = 1\\
         &\hspace{0.4cm}
         \left\{
         \begin{aligned}
            &\text{T$_2^\star$ : asymptotic bounded orbit }(\rho_1\leqslant \rho\leqslant \rho_C) \\
            &\text{T$_3^\star$ : asymptotic unbounded orbit }(\rho\geqslant \rho_C)\\
            &\text{T$^\star$ : circular unstable orbit }(\rho=\rho_C).
         \end{aligned}
         \right.
         &\\
      \end{aligned}
   \end{equation}
   We note that $\rho_2$ and $\rho_3$ are only defined if $\rho_C<1$. Figure~\ref{fig:phase} in the Appendix provides details of the phase portrait of the system, where the different types of trajectories can be easily identified (graph $n=2$). The extreme values of $\rho$ reachable by the particle (i.e. $\dot{\rho}=0$) can be obtained from Eq.~\eqref{eq:poteff} by solving the equation $U(\rho)=1$. This equation can be rewritten as
   \begin{equation}
      \rho\exp(\pm\rho) = \rho_0\,.
   \end{equation}
   The extreme values of $\rho$ reachable by the particle in the different cases are thus
   \begin{equation}\label{eq:r1r2r3}
      \rho_1 = W_0(\rho_0)
      \hspace{0.2cm};\hspace{0.2cm}
      \rho_2 = -W_0(-\rho_0)
      \hspace{0.2cm};\hspace{0.2cm}
      \rho_3 = -W_{-1}(-\rho_0),
   \end{equation}
   where $W_0$ and $W_{-1}$ are the Lambert functions. In accordance with Eq.~\eqref{eq:types}, $\rho_2$ and $\rho_3$ are only defined if $\rho_0<\exp(-1)$, that is, $\rho_C<1$.
   
   On the other hand, the conservation of $\rho_C$ (Eq.~\ref{eq:rho0}) allows us to write the angular velocity as a function of $\rho$ only (Fig.~\ref{fig:poteff}b). The stable equilibrium point at $\rho=\rho_0$ corresponds to $v=0$ (motionless particle), and the unstable equilibrium point at $\rho=\rho_C$ corresponds to a circular orbit with constant angular velocity. Whatever the trajectory, the angular velocity vanishes at $\rho=\rho_0$ and changes sign. The inner part of T$_1$ trajectories shows thus a unique loop away from the origin, whereas T$_2$ trajectories continuously rotate around the radius $\rho_0$. On the contrary, T$_3$ trajectories always rotate in the same direction around the origin (see Fig.~\ref{fig:poteff}c for some examples).
   
   A parametric expression of the trajectories can be easily obtained from the first integrals. Indeed, from Eqs.~\eqref{eq:rho0} and~\eqref{eq:poteff} we get
   \begin{equation}
      \begin{aligned}
         &\dot{\rho}^2 = \left(\frac{\mathrm{d}\rho}{\mathrm{d}\theta}\dot{\theta}\right)^2 = \left(\frac{\mathrm{d}\rho}{\mathrm{d}\theta}\frac{\ln(\rho/\rho_0)}{\rho^2}\right)^2  = 1 - \left(\frac{\ln(\rho/\rho_0)}{\rho}\right)^2 \\
         &\iff \mathrm{d}\theta^2 = \frac{\ln^2(\rho/\rho_0)}{\rho^4-\rho^2\ln^2(\rho/\rho_0)}\mathrm{d}\rho^2 ,
      \end{aligned}
   \end{equation}
   which gives
   \begin{equation}\label{eq:thetaR}
      \theta(\rho) = \theta_i\pm\int_{\rho_i}^{\rho}\varphi(\rho')\,\mathrm{d}\rho',
   \end{equation}
   with
   \begin{equation}\label{eq:phiR}
      \varphi(\rho) = \frac{\ln(\rho/\rho_0)}{\rho\sqrt{\rho^2-\ln^2(\rho/\rho_0)}},
   \end{equation}
   in which the initial conditions are written $(\rho_i,\theta_i)$. One can note that the integrand is singular in the extrema of $\rho$ (Eq.~\ref{eq:r1r2r3}), but the integral itself is always convergent (except for $\rho_C=1$, since in this case the particle makes an infinite number of loops before eventually reaching $\rho=\rho_C$). The $\pm$ sign in Eq.~\eqref{eq:thetaR} stands for the branches approaching $(-)$ and leaving $(+)$ the minimum radius. This definition by parts can be avoided by parametrising the trajectories by a parameter $s\in\mathbb{R}$. A possible parametrisation $\big(\theta(s),\rho(s)\big)$ of the three types of non-singular trajectories is given by
   \begin{equation}\label{eq:traj}
      \begin{aligned}
         &\theta(s) = \theta_i + \int_{s_i}^{s}\varphi\big(\rho(s')\big)\,\mathrm{d}s' \\
         &\text{and}\hspace{0.5cm}
         \left\{
         \begin{aligned}
            &\text{T$_1$: }\rho(s) = \rho_1 + |s| \\
            &\text{T$_2$: }\rho(s) = \rho_1 + \Big|\big((s-\Delta)\!\!\!\mod{2\Delta}\big) - \Delta\Big| \\
            &\text{T$_3$: }\rho(s) = \rho_3 + |s| \,,
         \end{aligned}
         \right.
      \end{aligned}
   \end{equation}
   where $\Delta = \rho_2-\rho_1$. With this parametrisation, $s(\tau)$ increases with $\tau$ and $\rho(s)$ is minimum at $s=0$.
   
   \subsection{Time information}\label{ssec:time}
   
   \begin{figure*}
      \centering
      \includegraphics[width=0.8\textwidth]{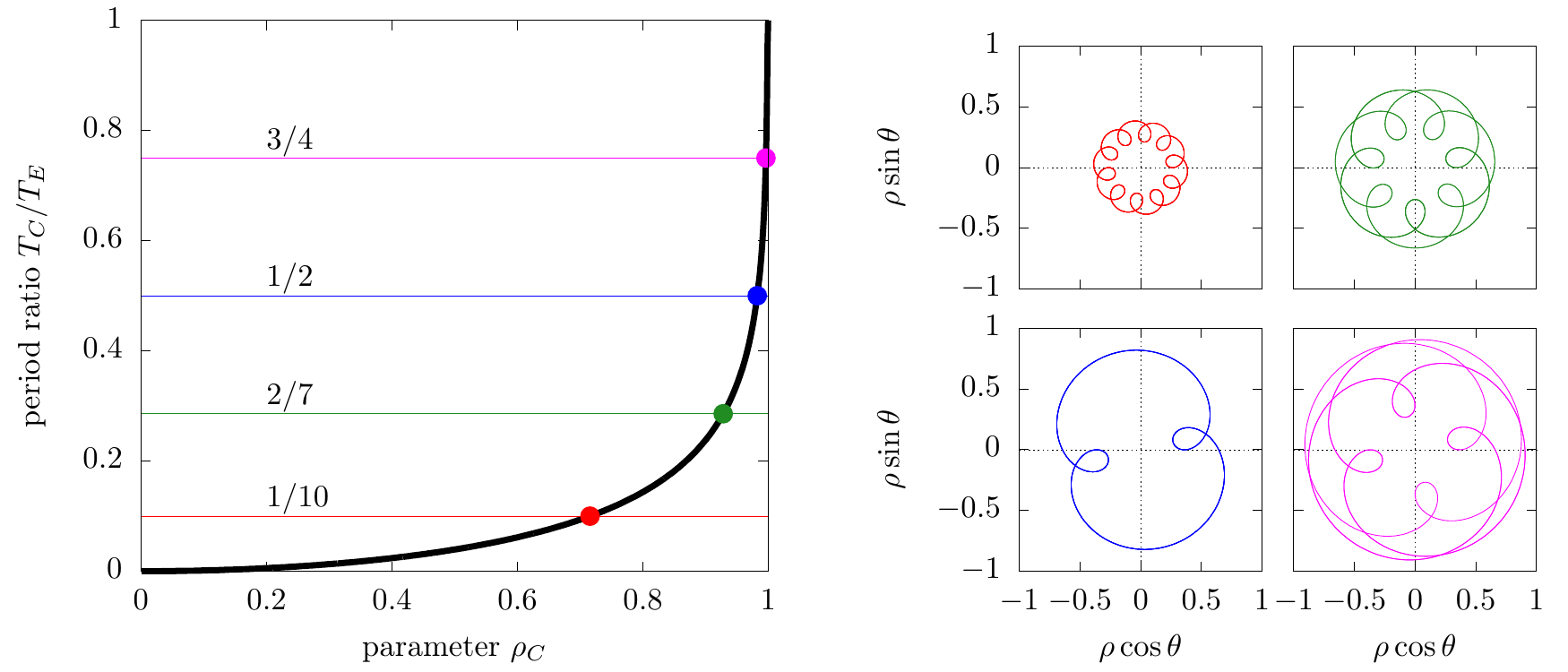}
      \caption{Left: Ratio of periods $T_C/T_E$ as a function of the parameter $\rho_C$, computed from Eq.~\eqref{eq:TC}. Some examples of rational values, corresponding to periodic orbits, are plotted as horizontal lines. Right: Same periodic orbits plotted in the physical plane. The corresponding parameter $\rho_C$ was obtained by a Newton method applied to Eq.~\eqref{eq:TC}.}
      \label{fig:perio}
   \end{figure*}
   
   By expressing the term $\rho^2\dot{\theta}^2$ from the energy constant (Eq.~\ref{eq:energy}) and by injecting it in the first equation of motion (Eq.~\ref{eq:motion1}), we get
   \begin{equation}
      \rho\ddot{\rho} + \dot{\rho}^2 - 1 = -\dot{\theta},
   \end{equation}
   which can be directly integrated to give
   \begin{equation}\label{eq:tau}
      \theta(\tau) = \tau - \rho\dot{\rho} + const.
   \end{equation}
   The polar angle is thus composed of a linear part plus a term proportional to $\dot{\rho}$. The physical meaning of the frequency $\omega_E$ (Eq.~\ref{eq:norm}) is now clear: it is the drift angular velocity of every particle. This is of particular interest for bounded trajectories. Indeed, they are quasi-periodic, with two proper frequencies: the ``drift'' frequency (rotation around the origin) and the ``loop'' frequency (small loop around the $\rho_0$ radius). Since $\dot{\rho}$ vanishes at the extreme values of $\rho$ (Eq.~\ref{eq:r1r2r3}), the period $T_C$ of the loops is
   \begin{equation}\label{eq:TC}
      \frac{1}{2}T_C = \tau(\rho_2) - \tau(\rho_1) = \theta(\rho_2) - \theta(\rho_1),
   \end{equation}
   while the period of the overall rotation around the origin is simply $T_E =2\pi$ (that is $2\pi/\omega_E$ in dimensional coordinates). Bounded trajectories represented in a frame rotating with $\tau$ consist thus only in the small loop around $\rho_0$. Periodic orbits are produced when the fraction $T_C/T_E$ is a rational number. Figure~\ref{fig:perio} shows the behaviour of $T_C/T_E$ as a function of the parameter $\rho_C$, along with some examples of periodic trajectories. We note that the two frequencies tends to be equal at $\rho_C$, that is, for the circular unstable trajectory (for which $\dot{\rho}=0$ at all time).
   
   More generally, Eq.~\eqref{eq:tau} can be used to express the time as a function of $\rho$ just like $\theta$ in Eq.~\eqref{eq:thetaR}. Expressing $\rho\dot{\rho}$ from Eq.~\eqref{eq:poteff}, we get
   \begin{equation}\label{eq:thetatau}
      \begin{aligned}
         \tau(\rho) - \tau_i &= \theta(\rho)-\theta_i \\ &\pm \left(\sqrt{\rho^2-\ln^2(\rho/\rho_0)}-\sqrt{\rho_i^2-\ln^2(\rho_i/\rho_0)}\right),
      \end{aligned}
   \end{equation}
   where $\pm$ means $(-)$ when the particle gets closer to the origin, and $(+)$ when it goes back. As before, a parameter $s\in\mathbb{R}$ can be used to avoid this double definition:
   \begin{equation}\label{eq:tauR}
      \tau(s) = \tau_i + \int_{s_i}^{s}\phi\big(\rho(s')\big)\,\mathrm{d}s' ,
   \end{equation}
   with
   \begin{equation}
      \phi(\rho) = \frac{\rho}{\sqrt{\rho^2-\ln^2(\rho/\rho_0)}}.
   \end{equation}
   This equation could have been obtained also directly from the energy constant (see Appendix~\ref{sec:comp}). It can be added among the parametrisation given by Eq.~\eqref{eq:traj} in order to compute the time at every position. From Eq.~\eqref{eq:thetatau}, one can note that in order to compute $\theta$ and $\tau$ at a given value of $s$, it is enough to compute only one integral.
   
   \section{Application to an incoming flux of particles}\label{sec:flux}
   Our first motivation for studying the inverse-square-law magnetic field is the deflection of solar wind protons as a result of their interactions with a cometary-type atmosphere. At very large distances from the comet, they can be considered as following parallel trajectories. In this section, we thus consider a permanent flux of particles initially evolving on parallel trajectories. As before, the $z$-component of the dynamics is trivial. We choose the orientation of the reference frame such that the initial velocity of the particles projected in the $(x,y)$ plane is along the $x$-axis ($\dot{x}_i=-v$ with $v>0$). At the position $x_i=d$, the magnetic field from Eq.~\eqref{eq:B} is activated. We aim to determine how the particles are distributed in the plane $(x,y)$ in the permanent regime, and in particular when $d\rightarrow\infty$. For finite $d$ distances, as we will consider in a first step, one can think of a continuous source of particles with the shape of a vertical infinite ``wall''.
   
   A similar setup was studied numerically by \cite{SHAIKHISLAMOV_2015} in a dipole $1/r^3$ field, but with the addition of a magnetopause (the particles were launched from a curved line instead of a fixed horizontal distance). As we will see, the two situations create similar features.
   
   \subsection{The cavity}
\begin{figure}
      \centering
      \includegraphics[width=\columnwidth]{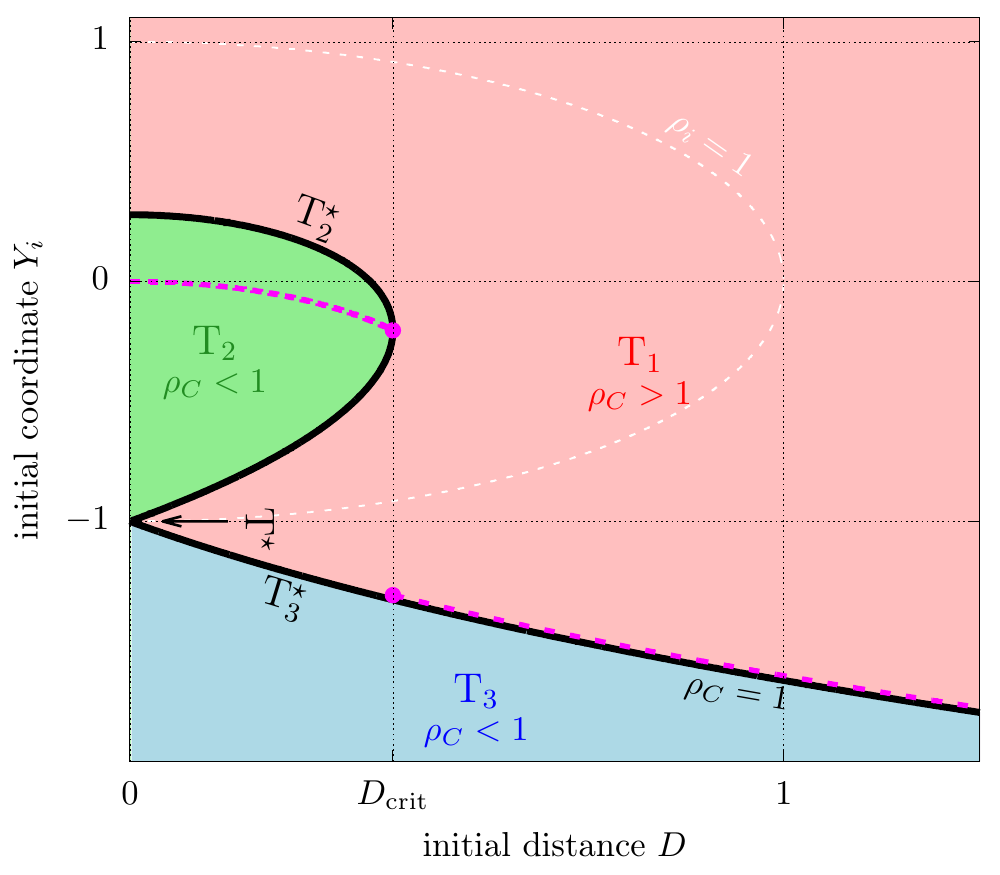}
      \caption{Type of orbit followed by a particle as a function of its initial position $(D,Y_i)$. The regions are coloured according to the value of $\rho_C(D,Y_i)$, and the level $\rho_C=1$ is represented by the black line. The types of orbits are labelled as in Eq.~\eqref{eq:types}. The magenta dashed line shows the initial position of the trajectory reaching the minimum radius over the whole vertical line. For $D<D_\mathrm{crit}$, it is a trajectory of type T$_2$. For $D>D_\mathrm{crit}$, it is a trajectory of type T$_1$ but infinitely close to the $\rho_C=1$ curve (see text).}
      \label{fig:zones}
   \end{figure}

   Since the particles have all the same velocity $v$, they have the same characteristic radius $r_E$ and drift frequency $\omega_E$ given by Eq.~\eqref{eq:norm}. We are thus able to use the normalised variables $\rho=r/r_E$ and $\tau=\omega_E\,t$ (same as previous section) in order to describe their motions in a common way. However, the characteristic radius $\rho_C$ of each particle is a function of its initial position along the $Oy$ axis. Using the normalised coordinates $Y_i=y_i/r_E$ and $D=d/r_E$, we get from Eq.~\eqref{eq:rho0}
   \begin{equation}\label{eq:rCy}
      \rho_C(Y_i) = \sqrt{D^2+Y_i^2}\exp\big(\mathrm{sgn}(k)\,Y_i+1\big),
   \end{equation}
   and thus $\rho_0(Y_i) \equiv \rho_C(Y_i)\exp(-1)$. The problem is about determining the different types of orbits followed by the particles as a function of $D$ and of their initial position $Y_i$. In the following, we suppose that $k$ is positive.\footnote{Since $\rho_C(Y_i)\big|_{-k} = \rho_C(-Y_i)\big|_k$, it is enough to study the case $k>0$. The case $k<0$ is obtained by mirror symmetry $Y_i\rightarrow-Y_i$.} First of all, we note that
   \begin{equation}\label{eq:lim}
      \lim\limits_{Y_i\rightarrow -\infty} \rho_C(Y_i) = 0
      \hspace{0.5cm}\text{and}\hspace{0.5cm}
      \lim\limits_{Y_i\rightarrow +\infty} \rho_C(Y_i) = \infty.
   \end{equation}
   The particles have thus all the possible values of $\rho_C$, including the critical one $\rho_C=1$ (Eq.~\ref{eq:types}). Let us write
   \begin{equation}\label{eq:Dlim}
      D_\mathrm{lim} = \frac{1}{2}\,,
   \end{equation}
   the limiting distance above which $\rho_C(Y_i)$ is monotonous. For $D>D_\mathrm{lim}$, there is thus only one trajectory with $\rho_C=1$ among the initial positions $Y_i$. For $D<D_\mathrm{lim}$, on the contrary, $\rho_C(Y_i)$ has a local maximum (larger than $1$) and a local minimum. It is straightforward to show that there is a critical distance,
   \begin{equation}\label{eq:Dcrit}
      \begin{aligned}
         D_\mathrm{crit} &= \frac{1}{2}\sqrt{-\,W_0(-2\exp[-2])\Big(W_0(-2\exp[-2])+2\Big)} \\
         &= 0.4023711712747059... \,,
      \end{aligned}
   \end{equation}
   such that if $D<D_\mathrm{crit}$, the local minimum of $\rho_C(Y_i)$ is smaller than $1$. This produces two other critical trajectories with $\rho_C=1$ (or only one in the limiting case $D=D_\mathrm{crit}$). In Appendix~\ref{sec:add}, Fig.~\ref{fig:rhoChar} shows the behaviour of all the characteristic lengths as a function of $D$, where the meaning of $D_\mathrm{lim}$ and $D_\mathrm{crit}$ is obvious. As a summary, the types of orbits followed by the particles are colour-coded in Fig.~\ref{fig:zones} as a function of their initial position. Trajectories of types T$_2$ and T$_3$ can be distinguished by considering their initial radius $\rho_i=\sqrt{D^2+Y_i^2}$, which should be smaller or larger than~$1$, respectively (since $\rho_2<1$ and $\rho_3>1$).
   
   In order to determine the distribution of the particles in the plane, useful information is given by the extreme radii reached by the particles. Each of them can be expressed in terms of $D$ and $Y_i$ by using Eqs.~\eqref{eq:r1r2r3} and~\eqref{eq:rCy}. For a fixed distance $D$, the minimum radius reached by the whole flux of particles is given by the minimum of $\rho_1(Y_i)$ over the T$_1$ and T$_2$ sets of orbits (red and green regions of Fig.~\ref{fig:zones}). For $D>D_\mathrm{crit}$, it corresponds to the inner loop of the asymptotic trajectory, that is, at the very limit between the red and blue zones of Fig.~\ref{fig:zones}. On the contrary, for $D<D_\mathrm{crit}$, the minimum value of $\rho_1$ is reached in the T$_2$ zone. The value of the minimum is
   \begin{equation}\label{eq:rhomin}
      \rho_\mathrm{cav} = 
      \left\{
      \begin{aligned}
         & W_0\Big(\sqrt{\gamma}\exp(-\gamma)\Big)
         &\text{for } D\leqslant D_\mathrm{crit} \\
         & W_0(\exp[-1]) = 0.278464542761...
         &\text{for } D\geqslant D_\mathrm{crit} \,,
      \end{aligned}
      \right.
   \end{equation}
   where $\gamma = \frac{1}{2}\left(1-\sqrt{1-4D^2}\right)$. Interestingly, for $D\geqslant D_\mathrm{crit}$, the value of $\rho_\mathrm{cav}$ is independent of the starting distance $D$. Moreover, we note that whatever the value of $D>0$, the minimum distance $\rho_\mathrm{cav}$ is never zero. This implies that among the whole flux of particles, none reaches the origin. In other words, the magnetic field naturally creates a cavity around the origin, devoid of any particle. The shape of this cavity can be inferred as follows:
   \begin{itemize}
      \item[$\bullet$] For $D<D_\mathrm{crit}$, the minimal radius is reached by a bounded orbit of type T$_2$. Hence, particles following this orbit come back periodically in $\rho_\mathrm{cav}$ and spread at all $\theta$ values (if we exclude periodic orbits as in Fig.~\ref{fig:perio}). The cavity in the permanent regime is thus circular.
      \item[$\bullet$] For $D\geqslant D_\mathrm{crit}$, the minimal radius is reached by an unbounded orbit of type T$_1$ for which $\rho_C\rightarrow 1$. We note that a particle following the exact critical trajectory $(\rho_C=1)$ never reaches the minimum radius, because it would have to pass through the asymptotic circular orbit (around which it circles infinitely, see Fig.~\ref{fig:poteff}a). However, particles starting from a position $Y_i$ slightly larger than the critical one do reach their minimal radii (though slightly larger than $\rho_\mathrm{cav}$) in a finite time. Moreover, these ``neighbour'' trajectories reach the latter with a different phase~$\theta$: the so-formed cavity is thus (asymptotically) circular with radius $\rho_\mathrm{cav}$.
   \end{itemize}
   Some examples of trajectories are presented in Appendix~\ref{sec:add} (Figs.~\ref{fig:cav1} and \ref{fig:cav3}), showing the formation of the cavity in the two regimes. We insist on the fact that it is circular in both cases, contrary to what was primarily suggested by \cite{BEHAR-etal_2017}. As we will see in the next section, the fact that it could seem elongated in the case $D>D_\mathrm{crit}$ comes from important contrasts in particle densities.
   
   In the case of solar wind protons deflected around an active comet, the starting distance can be considered as infinite ($D\gg D_\mathrm{crit}$). Hence, we need to verify that the trajectories produced by this simple model of the solar wind have a well-defined limit when $D\rightarrow\infty$. The function $\rho_C(Y_i)$ presented in Eq.~\eqref{eq:rCy} being monotonous whenever $D>D_\mathrm{lim}$ (Eq.~\ref{eq:Dlim}) and spanning all the possible values (as shown by Eq.~\ref{eq:lim}), the particles can be indifferently parametrised by their initial condition $Y_i$ or by their characteristic radius $\rho_C$. For a fixed value of $\rho_C$, the ratio $Y_i/D$ tends to $0$ when $D\rightarrow\infty$. This means that the initial angle $\theta$ of particles coming from infinity is~$0$. From Eq.~\eqref{eq:traj}, the expression of the trajectories is thus
   \begin{equation}\label{eq:trajinf}
      \theta(s) = \int_{+\infty}^{s}\varphi\big(\rho(s')\big)\,\mathrm{d}s'
      \hspace{0.3cm}\text{and}\hspace{0.3cm}
      \left\{
      \begin{aligned}
         &\text{T$_1$: }\rho(s) = \rho_1 + |s| \\
         &\text{T$_3$: }\rho(s) = \rho_3 + |s| \,,
      \end{aligned}
      \right.
   \end{equation}
   where the function $\varphi(\rho)$ is defined in Eq.~\eqref{eq:phiR}. This improper integral being convergent, the trajectories parametrised by their $\rho_C$ constant have indeed a well-defined limit when $D\rightarrow\infty$.
   
   It should be noted, though, that their initial position $Y_i$ tends to $-\infty$ (even if the ratio $D/Y_i$ tends to zero). The notion of ``impact parameter'' has thus no physical meaning in this problem. This information is crucial when dealing with simulations based on more realistic models of the solar wind because they are necessarily performed in a limited region of space (that is, for a finite value of $D$ and a finite range of $Y_i$). For now, we already know that the size of the simulation cells (considering a regular grid) should not exceed the radius of the cavity, which is the smallest scale of the system. As we will see in the next section, our simplistic model can also be used to infer the size of the simulation box required to obtain relevant results.
   
   \subsection{The caustic}
   
   In this section, we are interested in the relative density of particles in the $(x,y)$ plane in the permanent regime. As for the geometry of the trajectories (see Eq.~\ref{eq:trajinf} and text above), we should first determine if the density of particles in the $(X,Y)$ plane has a well-defined limit for $D\rightarrow\infty$. For a fixed value of $\rho_C$, we saw that $Y_i/D\rightarrow 0$ when $D\rightarrow\infty$, that is, $Y_i$ becomes negligible compared to $D$. The function $\rho_C(Y_i)$ from Eq.~\eqref{eq:rCy} behaves thus like $D\exp(Y_i+1)$, so we can replace the uniform distribution of the particles along the $Y_i$ axis by a uniform distribution of $\ln(\rho_C)$. Since, as shown above, the geometry of the trajectory for a given $\rho_C$ has itself a unique limit (Eq.~\ref{eq:trajinf}), the density map has also a well-defined limit for $D\rightarrow\infty$.

\begin{figure}
      \centering
      \includegraphics[width=0.8\columnwidth]{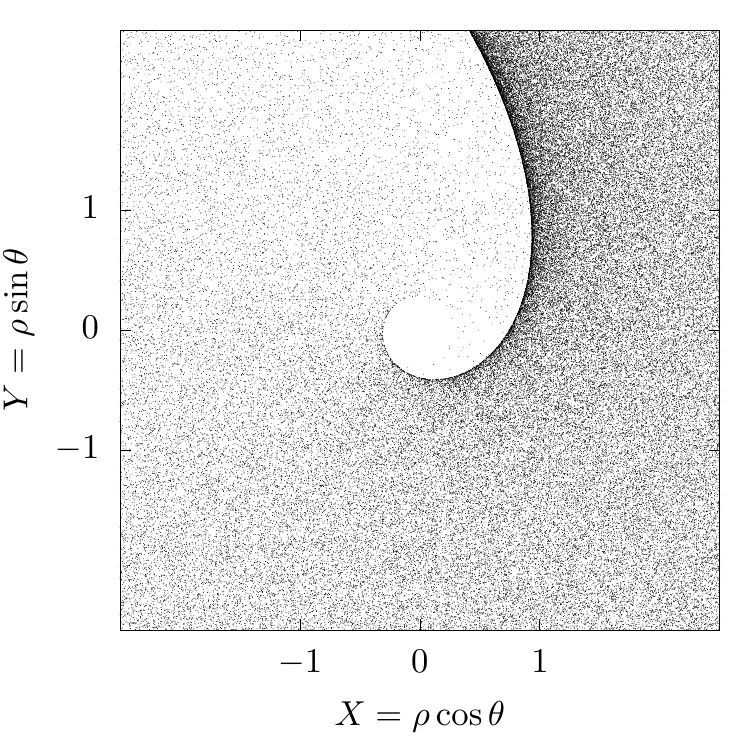}
      \includegraphics[width=0.8\columnwidth]{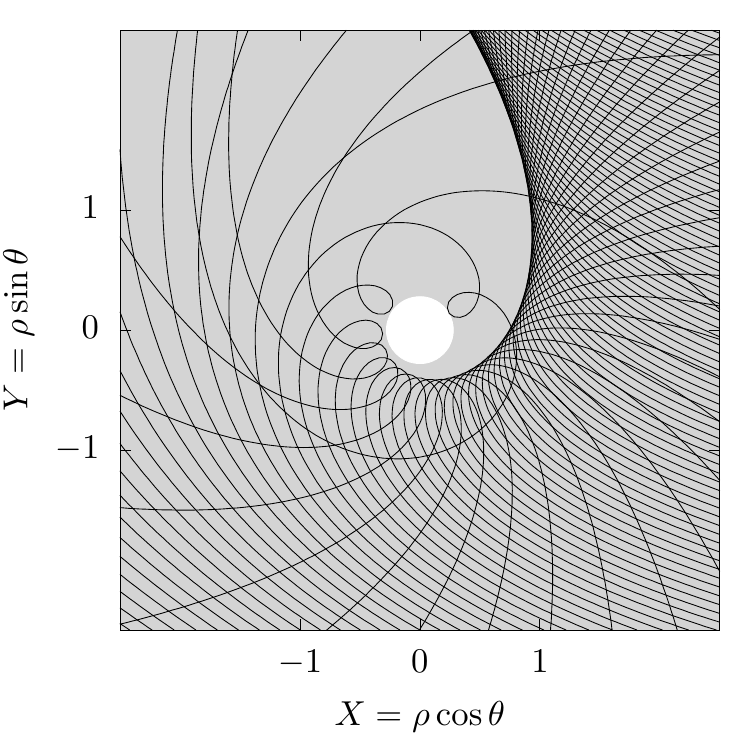}
      \caption{Top: Simulated density map of the particles around the origin for $D\rightarrow\infty$ (the particles come from the right). The inner cavity of radius $\rho_\mathrm{cav} = W_0(\exp[-1])$ is visible, as well as a caustic (line of overdensity). Bottom: Some trajectories evenly sampled along $\ln(\rho_C)$ are shown. The cavity is represented by the white disc.}
      \label{fig:dens}
   \end{figure}

   The relative density of particles can be simulated by distributing points randomly along trajectories evenly sampled along $Y_i$ (for finite $D$) or evenly sampled along $\ln(\rho_C)$ (for infinite $D$). Since the particles have all the same velocity, we must use an homogeneous distribution in time $\tau$. An illustration for infinite starting distance is given in Fig.~\ref{fig:dens}. As already reported by~\cite{SHAIKHISLAMOV_2015} for the $1/r^3$ magnetic field, a line of overdensity appears. This is a purely geometrical effect since, in the limit of this physical model, the particles do not interact which each other. This line is constituted of the points where two neighbouring trajectories cross each other. We will call it a ``caustic'' by analogy to light rays. For small values of $D$, several types of caustic appear. We will not go into details here, though, because small values of $D$ have no physical interest.
   
   \begin{figure*}
      \centering
      \includegraphics[width=0.8\textwidth]{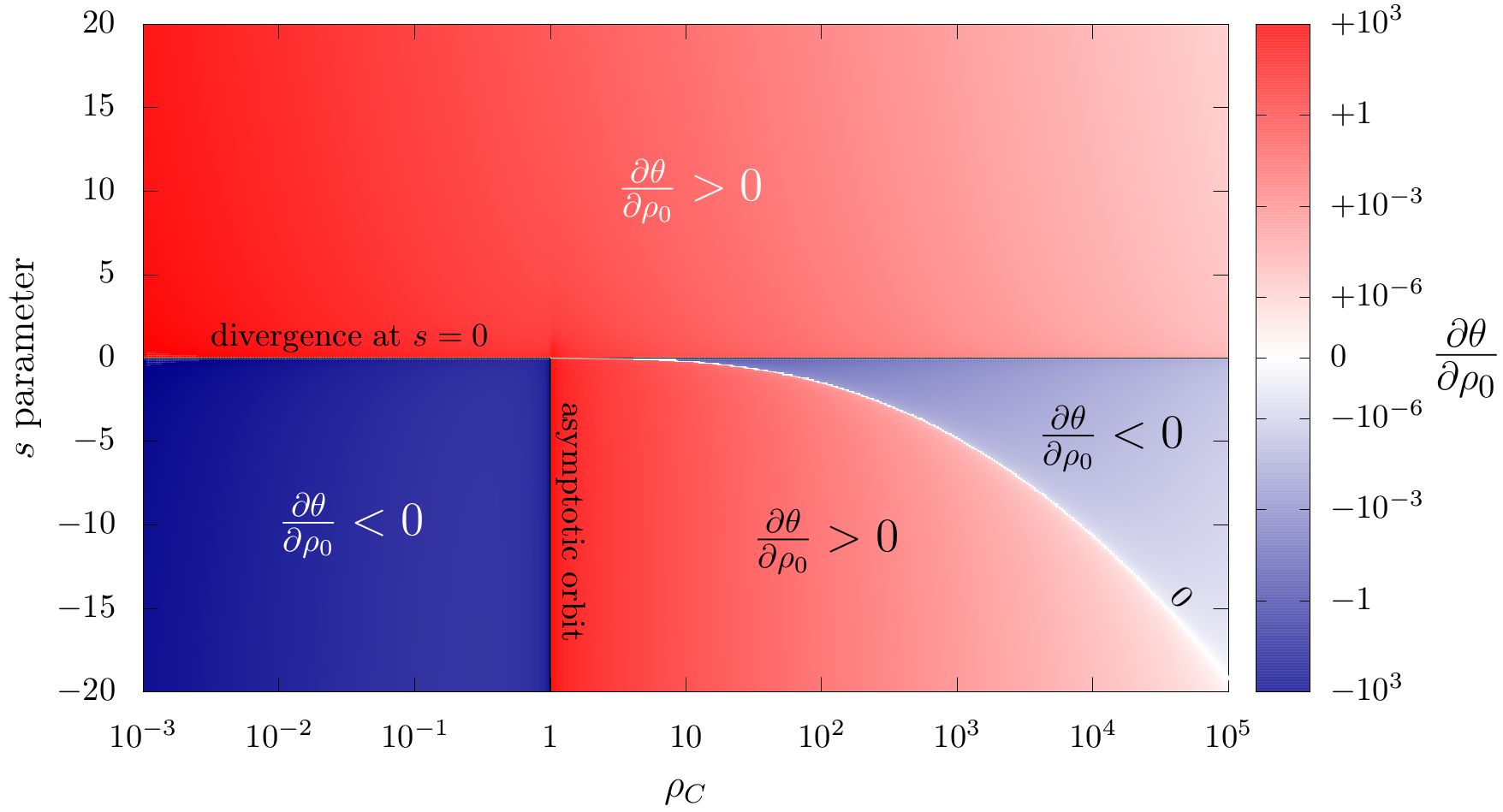}
      \caption{Value of $\partial\theta/\partial\rho_0$ in the $(\rho_C,s)$ plane for infinite $D$. Particles come from $s=\infty$, they reach their minimum radii at $s=0$, at which the last term of Eq.~\eqref{eq:dthdr0} diverges, and they go on with negative~$s$. The white line shows the level curve $\partial\theta/\partial\rho_0=0$, corresponding to the caustic (overdensity of particles). It is formed by the set of T$_1$ trajectories ($\rho_C>1$). See Fig.~\ref{fig:causticZoom} for its shape in the physical plane.}
      \label{fig:dth}
   \end{figure*}
   
   \begin{figure*}
      \centering
      \includegraphics[width=0.325\textwidth]{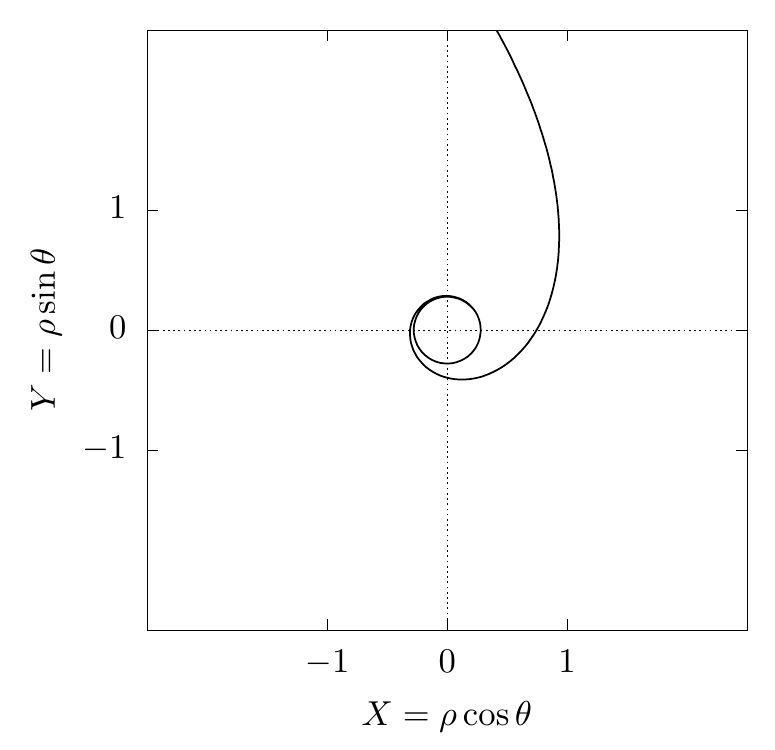}
      \includegraphics[width=0.325\textwidth]{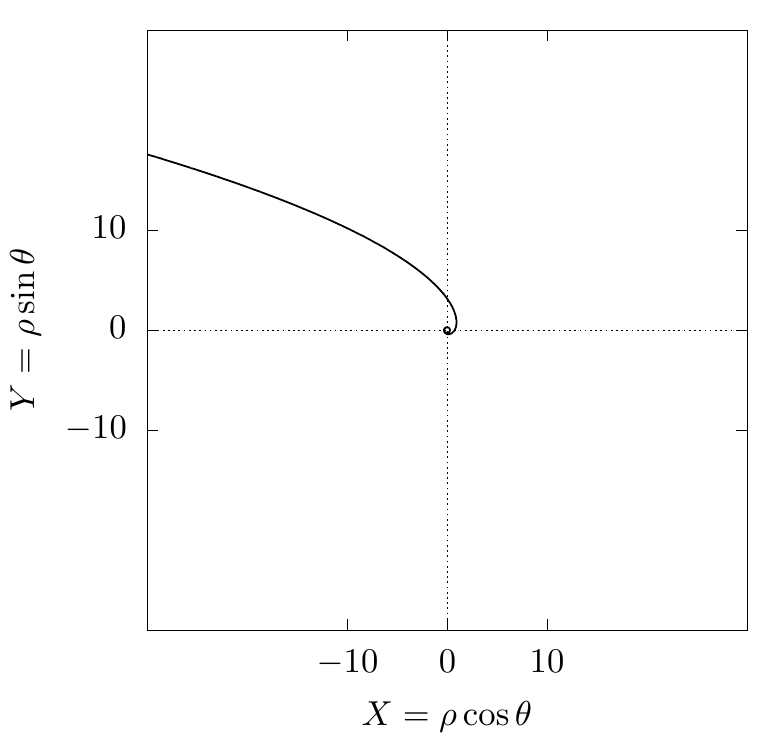}
      \includegraphics[width=0.325\textwidth]{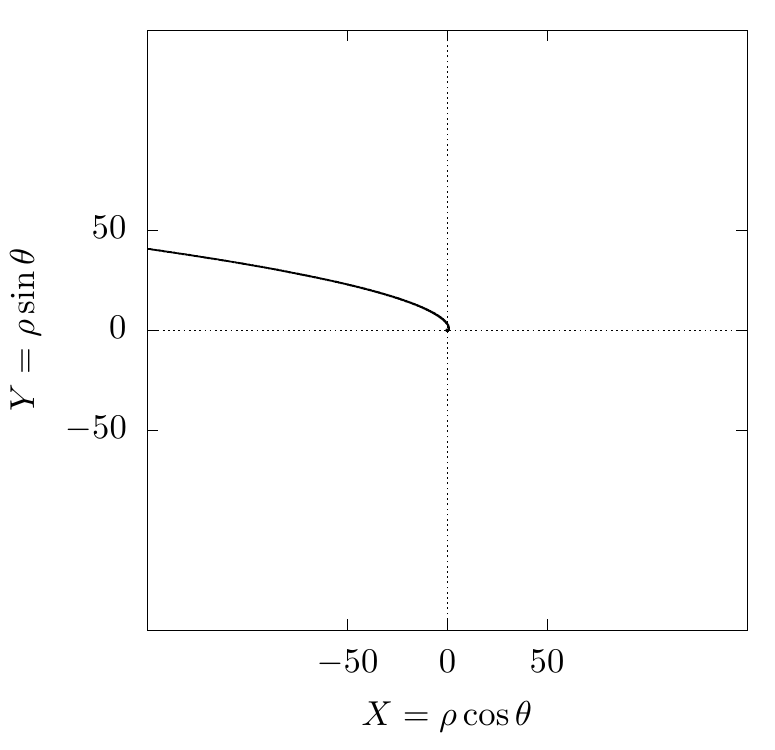}
      \caption{Form of the caustic obtained numerically by finding the root of $\partial\theta/\partial\rho_0$ (Eq.~\ref{eq:dthdr0}). Three different zoom levels are used, which can be interpreted as three level of cometary activity. The top panel presents the same scale as Fig.~\ref{fig:dens}, in which the density structure is clearly visible.}
      \label{fig:causticZoom}
   \end{figure*}
   
   In a general way, an overdensity region appears whenever the flux of particles is contracted, that is, when two trajectories of neighbouring initial conditions get closer to each other. This is quantified by the so-called variational equations. Let us consider a smooth function $f$ of time $t$, depending on one parameter $\alpha\in\mathbb{R}$ (which can be the initial condition $f(t=0)$). At a given time $t$, the distance $\mathrm{d}f$ between two curves with neighbouring values of the parameter $\alpha$ is at first order
   \begin{equation}
      \mathrm{d}f(\alpha;t) = \frac{\partial f}{\partial\alpha}(\alpha;t)\,\mathrm{d}\alpha
   \end{equation}
   \citep[see][for thorough details in the context of error propagations]{MILANI-GRONCHI_2010}. Of course, the distance between the two curves vanishes if they cross, implying that $\partial f/\partial\alpha=0$. In our case, the $\theta$ angle (Eq.~\ref{eq:trajinf}) plays the part of $f$, the radial variable $\rho$ plays the part of $t$, and the parameter $\rho_0$, itself bijectively linked to the initial condition $Y_i$, plays the part of $\alpha$. The variational equation can thus be written as
   \begin{equation}
      \mathrm{d}\theta(\rho_0;\rho) = \frac{\partial\theta}{\partial\rho_0}(\rho_0;\rho)\,\mathrm{d}\rho_0 \,.
   \end{equation}
   Using the chain rule, this partial derivative can be computed from Eq.~\eqref{eq:trajinf}, considering $s$ as a function of $\rho$, itself a function of $\rho_0$ via $\rho_1$ or $\rho_3$. We obtain
   \begin{equation}\label{eq:dthdr0}
      \begin{aligned}
         \frac{\partial\theta}{\partial\rho_0} &= \int_{+\infty}^{s}\left(\frac{\partial\varphi}{\partial\rho_0}\big(\rho(s')\big) + \frac{\partial\rho}{\partial\rho_0}\,\frac{\partial\varphi}{\partial\rho}\big(\rho(s')\big)\right)\mathrm{d}s' \\
         &- \mathrm{sgn}(s)\frac{\partial\rho}{\partial\rho_0}\varphi\big(\rho(s)\big) \,,
      \end{aligned}
   \end{equation}
   with
   \begin{equation}
      \begin{aligned}
         &\frac{\partial\varphi}{\partial\rho_0} = \frac{-\rho/\rho_0}{\left[\rho^2-\ln^2(\rho/\rho_0)\right]^{3/2}} \,, \\
         &\frac{\partial\varphi}{\partial\rho} = \frac{\ln^3(\rho/\rho_0) + \rho^2\left[1-2\ln(\rho/\rho_0)\right]}{\rho^2\left[\rho^2-\ln^2(\rho/\rho_0)\right]^{3/2}} \,,
      \end{aligned}
   \end{equation}
   and
   \begin{equation}
      \frac{\partial\rho}{\partial\rho_0} = 
      \left\{
      \begin{aligned}
         \frac{\partial\rho_1}{\partial\rho_0} &= \frac{1}{\rho_0}\,\frac{\rho_1}{1+\rho_1}
         \hspace{0.5cm}\text{for trajectories of type T}_1 \,, \\
         \frac{\partial\rho_3}{\partial\rho_0} &= \frac{1}{\rho_0}\,\frac{\rho_3}{1-\rho_3}
         \hspace{0.5cm}\text{for trajectories of type T}_3 \,.
      \end{aligned}
      \right.
   \end{equation}
   For finite values of $D$, an analogous formula can be obtained from Eq.~\eqref{eq:traj}, containing additional terms due to the initial conditions. Figure~\ref{fig:dth} shows the general form of $\partial\theta/\partial\rho_0$ in the $(\rho_C,s)$ plane. Particles with $\rho_C<1$ do not produce any accumulation (they rather spread). Particles with $\rho_C>1$, on the contrary, arrive at a point where $\partial\theta/\partial\rho_0$ becomes zero and changes sign. This means that neighbouring trajectories cross in this point, creating an overdensity. The curve along which $\partial\theta/\partial\rho_0$ is zero can be obtained numerically using a Newton-type algorithm. Its shape in the $(X,Y)$ plane is presented in Fig.~\ref{fig:causticZoom} (it should be compared to the density map of Fig.~\ref{fig:dens}). For particles coming from infinity, the shape of the caustic only depends on the characteristic radius $r_E$, which acts as a scaling parameter. For solar wind protons deflected around a comet, this means that whatever the cometary activity (expressed in the $k$ parameter), the structure formed by the proton trajectories is always exactly the same, though it is seen at a different ``zoom level''. This is illustrated in Fig.~\ref{fig:causticZoom}.
   
\begin{figure}
      \centering
      \includegraphics[width=\columnwidth]{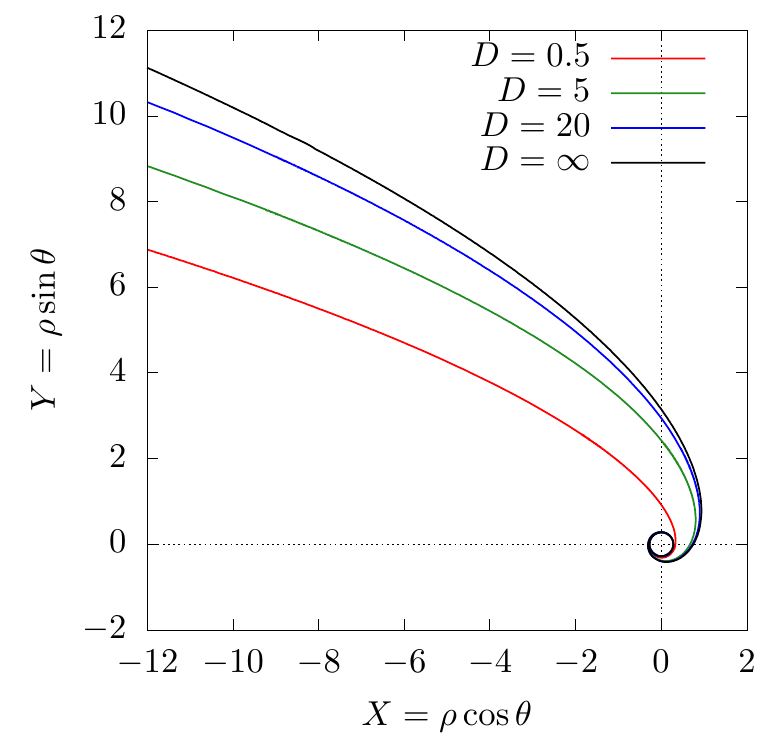}
      \caption{Form of the caustic for different starting distances $D \geqslant D_\mathrm{crit}$. The distortion caused by the use of a finite value of $D$ is shown by the difference with the $D=\infty$ curve (black line).}
      \label{fig:simubox}
   \end{figure}

   As mentioned earlier, complex numerical simulations of the interaction of solar protons with cometary ions are always limited to finite simulation boxes. In practice, this means that solar particles, supposed unaffected yet by the comet, are launched from a finite distance $D$. This necessarily distorts the dynamical structures, as already pointed out by \cite{KOENDERS_2013} for high-activity comets. In our case, by comparing the shape of the caustic for different starting distances $D$, our simplistic model can give an estimate of the error introduced by the finite-sized simulation boxes. This is shown in Fig.~\ref{fig:simubox}: simulations with a small box tend to underestimate the opening angle of the caustic. The error is thus larger at larger distances from the nucleus (but the cavity radius is unaffected as long as $D\geqslant D_\mathrm{crit}$).
   
   \section{Parameter values for a realistic comet}\label{sec:rosetta}
   From Eq.~\eqref{eq:rhomin}, we know that solar wind protons are in the regime for which the radius of the circular cavity is independent of $D$. Switching back to dimensional quantities, it writes $r_\mathrm{cav} \approx 0.28\,r_E$. The radius of the cavity depends thus only on $r_E = |k|/v$, that is, on the incident velocity of the particles and on the $k$ constant of the effective magnetic field. In particular, the cavity boundary was crossed by the \emph{Rosetta} spacecraft: knowing $v$, its distance from the comet at time of crossing allows us to measure the $k$ parameter (assuming that the solar wind protons did follow this simple model). Order-of-magnitude estimates can be obtained from Fig.~1 by \cite{BEHAR-etal_2017}: the spacecraft crossed the boundary from inside to outside the cavity in December 2015, when the comet was at about $1.7$~au from the Sun. The data give $v=300$~km/s and $r_\mathrm{cav}=130$~km at the time of crossing, resulting in a characteristic length $r_E\approx 470$~km and a $k$ parameter of the order of $10^5$~km$^2$/s.
   
   As shown in Sect.~\ref{ssec:model}, the value of $k$ is proportional to the outgassing rate $Q$ of the comet \citep[see the companion paper by][for details]{BEHAR-etal_2018a}. Actually, considering the very high velocity of the solar protons, the change of $k$ due to the varying cometary activity can safely be modelled as an adiabatic process. Each time of an observation by \emph{Rosetta} corresponds thus to a different value of $k$ (or equivalently $r_E$). Still assuming that the solar wind protons did follow the dynamics described in this paper, the parameter $k$ can be estimated at any time from the observed deflection of solar particles. Indeed, knowing the position of the spacecraft during each observation in the comet-Sun-electric frame (CSE), we just have to rescale the picture (that is, to find the unit length $r_E$), such that the incoming flux of particles is indeed deflected by the observed amount at this specific position. This method will be presented in detail in a forthcoming article, in which the data points will be systematically compared to theoretical values. It leads to the cavity radius being larger than $5$~km when the comet is closer than $2.6$~au from the Sun, and growing beyond $1500$~km at perihelion.
   
\section{Conclusion}
   During most of their trajectory around the Sun, comets are in a low-activity regime. When studying the dynamics of cometary and solar wind ions, this results in a gyration scale larger than the interaction region. In this situation, solar wind protons can be efficiently modelled by test particles subject to a magnetic-field-like force proportional to $1/r^2$ (in a cometocentric reference frame). In this article, we provided a full characterisation of their trajectories in the plane perpendicular to this field.
   
   As for every autonomous vector field with rotational symmetry, the system admits two conserved quantities: the kinetic energy $E$ and a generalised angular momentum $C$. In our case, both of them can be turned into characteristic radii $r_E$ and $r_C$, which entirely define the dynamics (throughout the text, we rather use the adimensional quantity $\rho_C=r_C/r_E$). There are three families of trajectories: two of them gather unbounded orbits ($r_C>r_E$ and $r_C<r_E$), and the other one contains quasi-periodic bounded orbits ($r_C<r_E$). A bifurcation occurs at $r_C=r_E$, with a homoclinic orbit asymptotic to a circle of radius $r_C$ (hyperbolic equilibrium point) and two branches coming from and going to infinity. Generic analytical expressions of the trajectories $(r,\theta,t)$ are obtained, of the form $\theta(r) = \omega_E\,t(r) + f(r)$, where $\omega_E$ is a constant, $f(r)$ is an explicit function, and the time $t(r)$ is defined by an integral.
   
   When considering an incoming flux of particles coming from infinity on parallel trajectories and at the same velocity, a cavity is naturally created around the origin. This cavity, entirely free of particle, is circular with radius $r_\mathrm{cav}\approx 0.28\,r_E$. Extending away from it, a curve of overdensity of particles spreads similarly to an optical caustic. This overdensity curve has no explicit expression but its shape in the plane can be computed at an arbitrary precision. The whole setting depends only on $r_E$, which acts as a scaling parameter.
   
   If we model the motion of solar wind protons around comet 67P by this simple dynamics, the radius $r_E$ can be calibrated from \emph{Rosetta} plasma observations. From the arrival of \emph{Rosetta} in the vicinity of the comet until the signal turn-off when reaching the cavity boundary, $r_E$ grew from a few kilometres up to about $470$~km. This gives not only a qualitative understanding of the observed deflection of solar particles and the formation of the cavity, but also the relevant scales for the problem. In particular, if refined simulations of solar wind are used, we stress the need for simulation boxes much larger than the characteristic radius $r_E$ (to account for the caustic shape) and a grid much finer than $r_E$ (if one wants to resolve the cavity structure).
   
   According to the results by \citet{BEHAR-etal_2018a}, it is important to note that the capacity of this simple dynamics to account for the motion of solar wind protons around a comet is increasing with the distance to the nucleus: the farther away from the nucleus, the better the model. In other words, physical assumptions on which the physical model is based may start to crumble at the origin (the nucleus) first, leaving the modelled deflection far from the nucleus unaffected.
   
   Comparisons to a generic magnetic field proportional to $1/r^n$, added in Appendix~\ref{sec:comp}, reveal similar features whenever $n>1$. Albeit the radius of the cavity and the precise shape of the caustic are different for each $n$, the specific choice of $1/r^2$, if only taken on empirical grounds, would be difficult to justify. However, this law can now be retrieved from the physical modelling of solar wind protons and cometary activity, as it is presented by \cite{BEHAR-etal_2018a,BEHAR-etal_2018b}.

\begin{acknowledgements}
   We thank the anonymous referee for her or his thorough reading, which led to a much better version of the article.
\end{acknowledgements}

\bibliographystyle{aa}
\bibliography{EBM2}

\appendix
\section{Complementary figures}\label{sec:add}
\begin{figure*}
      \centering
      \includegraphics[width=0.8\textwidth]{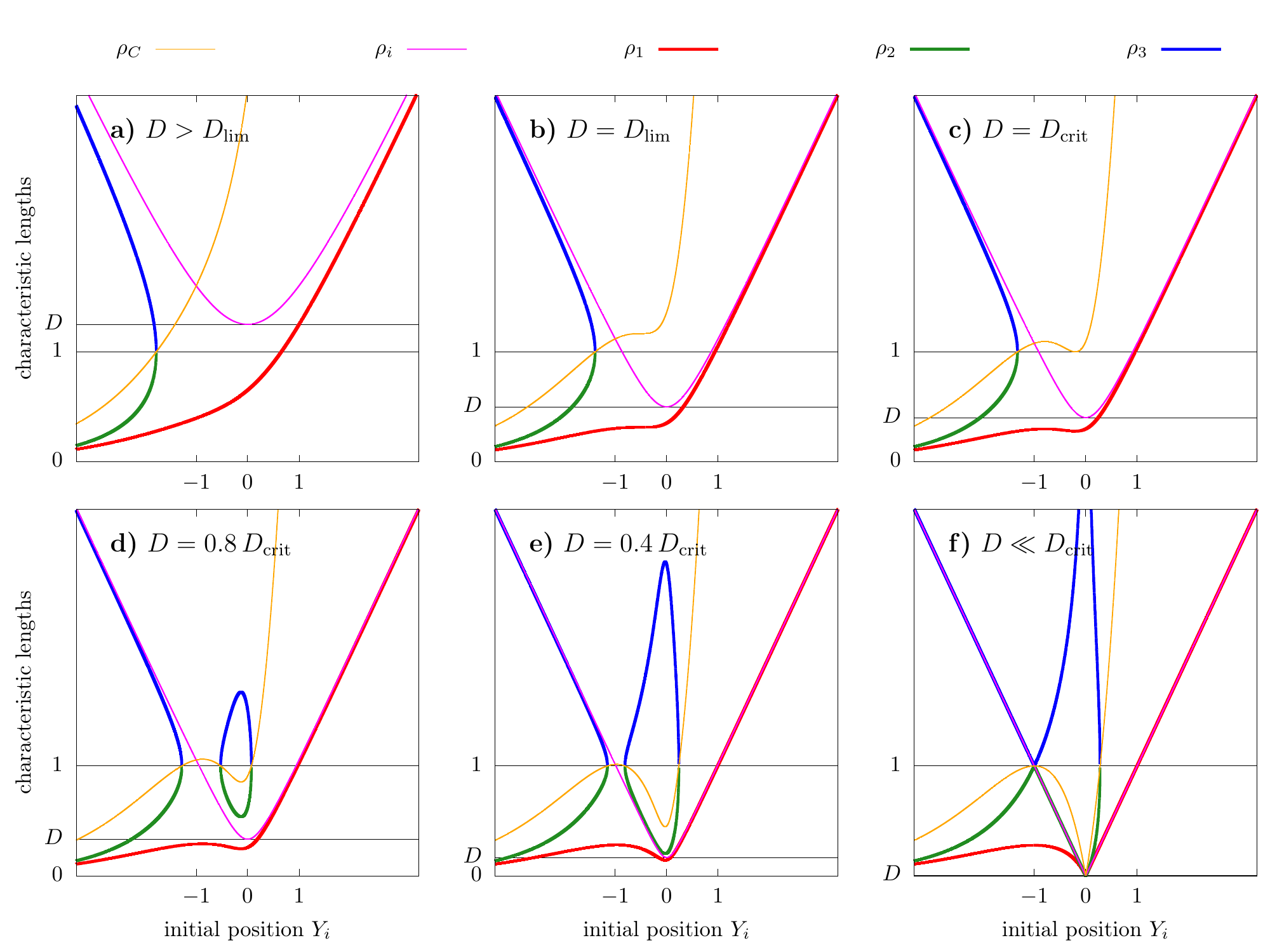}
      \caption{Characteristic lengths along the line of initial position $Y_i$ for different distances $D$. Each curve is labelled above the graphs: the parameter $\rho_C$ (Eq.~\ref{eq:rCy}) is drawn in yellow; the initial starting distance $\rho_i=\sqrt{D^2+Y_i^2}$ is drawn in magenta; the extreme reachable radii $\rho_1$, $\rho_2$, and $\rho_3$ (Eq.~\ref{eq:r1r2r3}) are drawn in red, green, and blue. As a function of $Y_i$, the parameter $\rho_C$ is monotonous if $D\geqslant D_\mathrm{lim}$ (\textbf{a}, \textbf{b}), it crosses $1$ in one additional point if $D=D_\mathrm{crit}$ (\textbf{c}), and in two additional points if $D<D_\mathrm{crit}$ (\textbf{d}, \textbf{e}, \textbf{f}). The type of trajectory of the particle with initial position $Y_i$ is determined by the location of its initial distance $\rho_i$ with respect to the characteristic lengths $\rho_1$, $\rho_2$, and $\rho_3$: the trajectory is of type T$_2$ when $\rho_1<\rho_i<\rho_2$; T$_3$ when $\rho_i>\rho_3$; and T$_1$ when $\rho_i>\rho_1$ (with no $\rho_2$, $\rho_3$). We note that trajectories of type T$_2$ are only possible for $D<D_\mathrm{crit}$ (\textbf{d}, \textbf{e}, \textbf{f}).}
      \label{fig:rhoChar}
   \end{figure*}

\begin{figure}
      \centering
      \includegraphics[width=0.6\columnwidth]{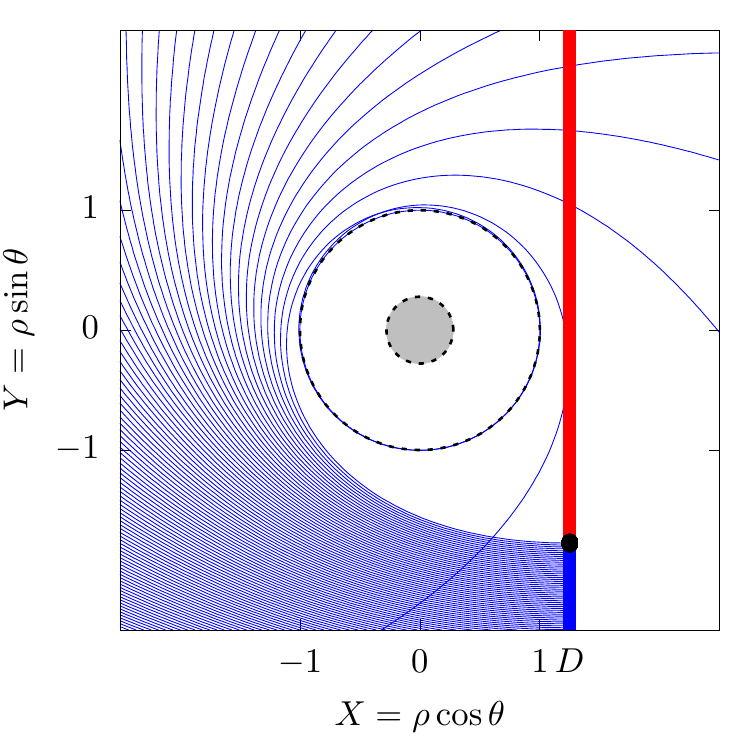}
      \includegraphics[width=0.6\columnwidth]{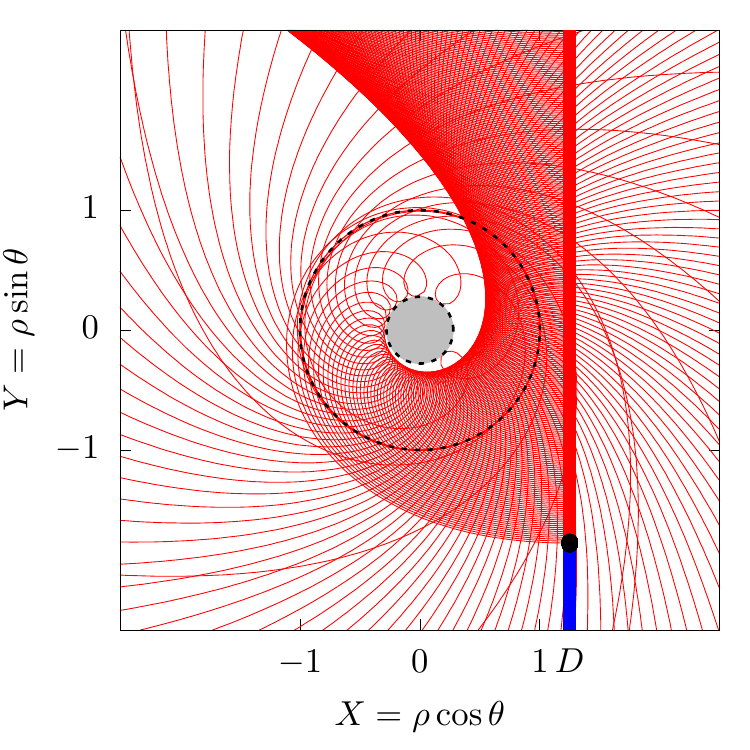}
      \caption{Flux of particles coming from the line $X=D$, with $D=1.25>D_\mathrm{crit}$. The two types of possible orbits are represented separately, with the same colour code as in Fig.~\ref{fig:zones}. Top: Unbounded trajectories of type T$_3$ are represented in blue. They approach a minimum distance equal to $1$ but never reach it exactly (outer dashed circle, around which they can perform an arbitrary number of turns before going back). In the limiting case where $\rho_C=1$ (black initial condition), the particle makes a infinite number of turns as $\rho\rightarrow 1$. Bottom: Unbounded trajectories of type T$_1$ are represented in red. They cross the characteristic radius $\rho_0$, at which their angular velocity is inverted (see the small loops). For initial positions $Y_i$ tending to the black point, the minimal distance reached by the particle tends to $\rho_\mathrm{cav}=W_0(\exp[-1])$ (inner dashed circle, see Eq.~\ref{eq:rhomin}). If we consider an infinite number of particles, this forms a cavity with radius $W_0(\exp[-1])$ (grey disc).}
      \label{fig:cav1}
   \end{figure}

\begin{figure}
      \centering
      \includegraphics[width=0.6\columnwidth]{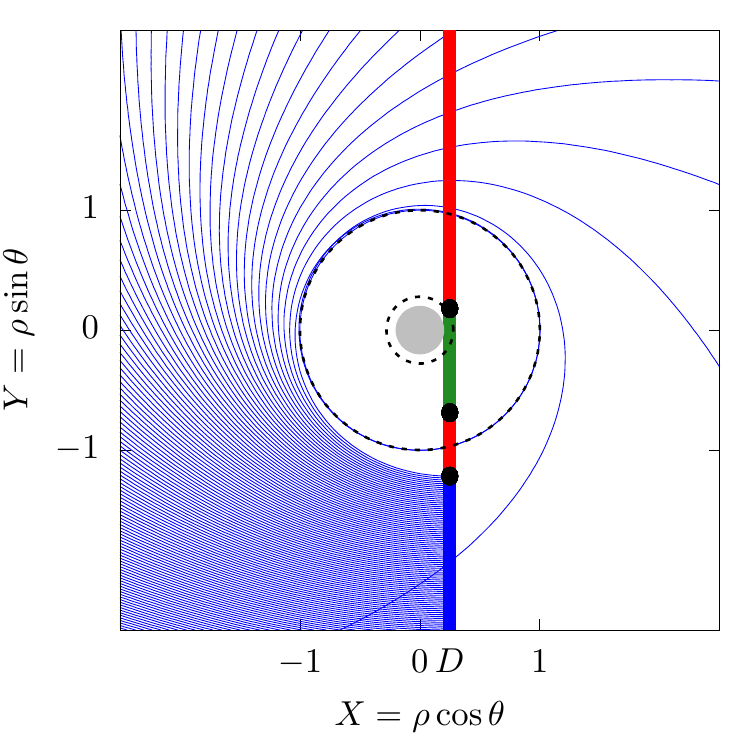}
      \includegraphics[width=0.6\columnwidth]{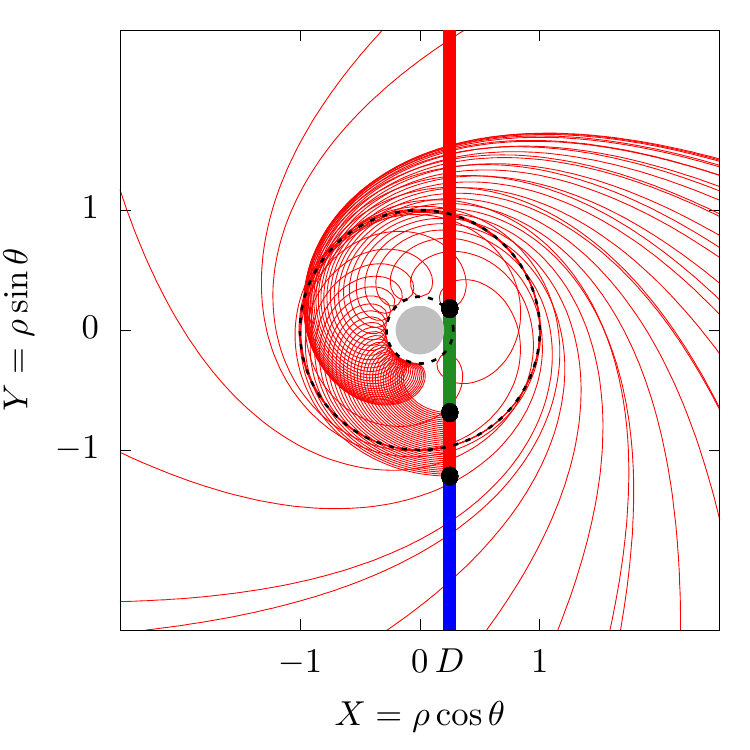}
      \includegraphics[width=0.6\columnwidth]{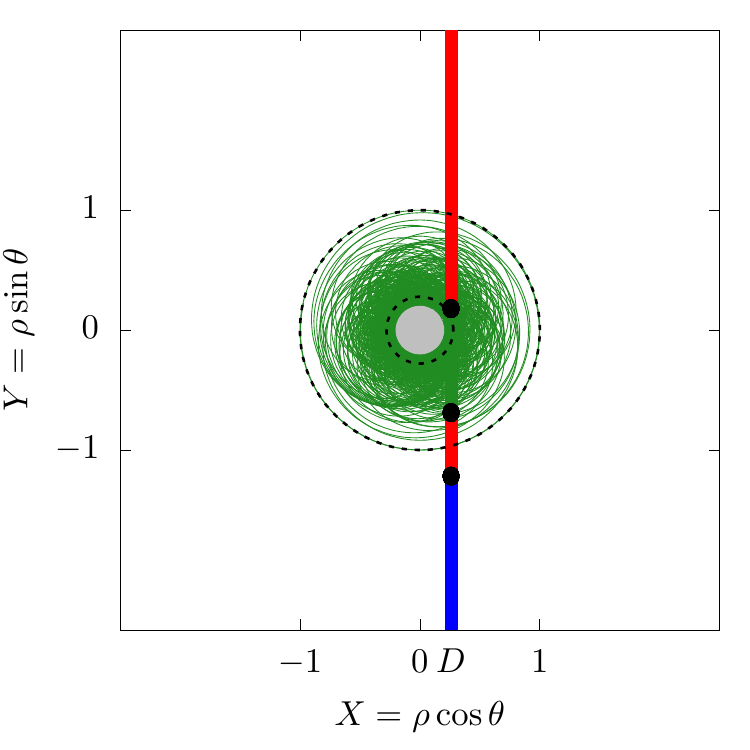}
      \includegraphics[width=0.6\columnwidth]{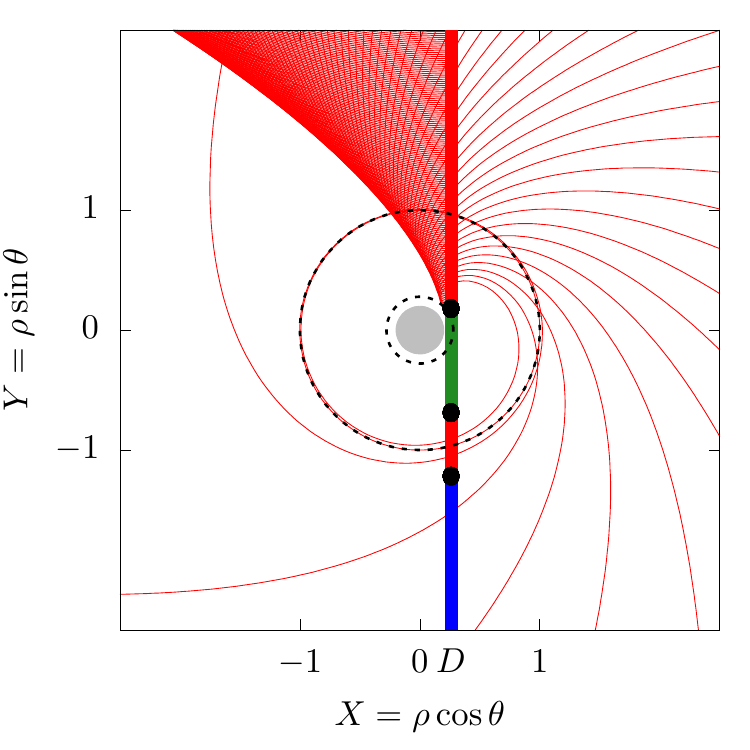}
      \caption{Same as Fig.~\ref{fig:cav1} but for $D=0.25<D_\mathrm{crit}$. An interval of initial conditions produces bounded orbits (in green), which loops forever inside the unit circle. They approach closer to the origin than the red trajectories, producing a smaller circular cavity (grey disc, see Eq.~\ref{eq:rhomin}). For comparison, the same circles as in Fig.~\ref{fig:cav1} are represented. We note that the red trajectories are still limited by the $W_0(\exp[-1])$ radius (inner dashed circle).}
      \label{fig:cav3}
   \end{figure}

   See Figs.~\ref{fig:rhoChar}, \ref{fig:cav1} and \ref{fig:cav3}.
   
\section{Comparison to other powers of $1/r$}\label{sec:comp}
   Starting from the pioneering work by \cite{STORMER_1907} applied to the geomagnetic field, there is a vast literature about the trajectories of particles in the equatorial plane of a magnetic dipole, for which the magnetic field is perpendicular to the plane and proportional to $1/r^3$. The present paper reveals that the $1/r^2$ field has numerous similarities, so we propose here to compare the dynamics driven by the different powers of $1/r$.
   
   Let us introduce a positive integer $n\in\mathbb{N}^*$, and a physical constant $k_n$ (in unit length to the power $n$ per unit time). The equations of motion in polar coordinates for a magnetic field proportional to $1/r^n$ are
   \begin{align}
      \ddot{r}-r\dot{\theta}^2 &= \frac{k_n}{r^{n-1}}\dot{\theta} \label{eq:gen1}\\
      r\ddot{\theta}+2\dot{r}\dot{\theta} &= -\frac{k_n}{r^n}\dot{r} \,.\label{eq:gen2}
   \end{align}
   As before, these equations imply the conservation of the velocity norm $v = \sqrt{\dot{r}^2 + r^2\dot{\theta}^2}$ and a generalised angular momentum $c_n$ obtained by direct integration of Eq.~\eqref{eq:gen2}.
   
   As for any planar problem with rotational symmetry, there exists an expression of the solutions $(\theta,t)$ as a function of $r$ defined by an integral. Indeed, the conservation of $c_n$ allows us to express $\dot{\theta}$ as a function or $r$, which can be injected in the velocity norm. The solution is finally obtained by quadrature (see Formulas~\ref{eq:traj} and~\ref{eq:tauR} obtained for $n=2$).
   
   Despite this general way of resolving the equations, the case $n=2$ is special. Indeed, it is the only one for which Eq.~\eqref{eq:gen1} is also directly integrable, by expressing the $\dot{\theta}^2$ term from the energy. This allowed us to express explicitly the time as a function of $\theta$ and $r$ (Eq.~\ref{eq:thetatau}). In practice, this means the case $n=2$ is the only one for which the ``drift'' proper frequency of all the trajectories is $\omega_E$ (see Sect.~\ref{ssec:time}). For any other power of $n$, the quantity $\omega_E$ is only the frequency of the unstable circular orbit (see below); the drift frequency of the other trajectories is a function of $c_n$ and thus different for all of them \citep{HAMLIN-etal_1961,AVRETT_1962}. This somewhat complicates the search for periodic trajectories \citep{GRAEF-KUSAKA_1938}.
   
   The case $n=1$ must also be taken separately, because since $k_1$ has the dimension of a velocity, the constant $v$ cannot be turned into a characteristic length analogous to $r_E$ (Eq.~\ref{eq:norm}). The dynamics can though be studied by using the ``effective potential'' method. When considering an incoming flux of particles as in Sect.~\ref{sec:flux}, we show that the launch distance $d$ becomes the scaling parameter of the system. There is thus no limit when $d\rightarrow\infty$. This means that the case $n=1$ has no physical meaning for this setting.
   
   For $n\geqslant 2$, the parameter $k_n$ naturally defines a characteristic length and a characteristic frequency:
   \begin{equation}\label{eq:rEgen}
      r_E = \left(\frac{|k|}{v}\right)^\frac{1}{n-1}
      \hspace{1cm};\hspace{1cm}
      \omega_E = -\frac{k}{r_E^n} \ .
   \end{equation}
   Examples for $n=2$ and $n=3$ can be found in Eq.~\eqref{eq:norm} and~\cite{STORMER_1930}. As shown in Sect.~\ref{sec:dyn}, they can be used to define dimensionless coordinates $\rho=r/r_E$ and $\mathrm{d}\tau=\omega_E\,\mathrm{d}t$. The case $n=2$ is studied in detail above, so we will now suppose that $n>2$. Using the dimensionless coordinates, the equations of motion and conserved quantities rewrite as
   \begin{equation}
      \left\{
      \begin{aligned}
         &\rho^{n-1}\ddot{\rho}-\rho^n\dot{\theta}^2 = -\dot{\theta} \\
         &\rho^2\ddot{\theta}+2\rho\dot{\rho}\dot{\theta} = \frac{\dot{\rho}}{\rho^{n-1}}
      \end{aligned}
      \right.
      \iff
      \left\{
      \begin{aligned}
         1 &= \dot{\rho}^2 + \rho^2\dot{\theta}^2 \\
         C_n &= \rho^2\dot{\theta} + \frac{1}{n-2}\frac{1}{\rho^{n-2}} \,,
      \end{aligned}
      \right.
   \end{equation}
   where, as before, the dot now means derivative with respect to the normalised time $\tau$. We note that $C_n$ is now the angular momentum at infinity, or equivalently, the impact parameter times the constant velocity norm. Using the ``effective potential'' method as in Eq.~\eqref{eq:poteff}, we get
   \begin{equation}
      1 = \dot{\rho}^2 + U_n(\rho)
      \hspace{0.3cm}\text{with}\hspace{0.3cm}
      U_n(\rho) = \left(\frac{(n-2)C_n\,\rho^{n-2}-1}{(n-2)\rho^{n-1}}\right)^2 \,.
   \end{equation}
   The $\rho_0$ and $\rho_C$-like characteristic lengths associated to this potential would be
   \begin{equation}
      \begin{aligned}
         \rho_0 &= \left(\frac{1}{(n-2)C_n}\right)^{\frac{1}{n-2}} \\
         \rho_C &= \left(\frac{n-1}{(n-2)C_n}\right)^{\frac{1}{n-2}} = (n-1)^{\frac{1}{n-2}}\,\rho_0\ ,
      \end{aligned}
   \end{equation}
   but since they become negative or complex numbers when $C_n<0$ (or even undefined when $C_n=0$), they would not have such a general physical meaning as for $n=2$. The system is thus better parametrised by $C_n$ itself, or by a wisely chosen parameter $\gamma_n$:
   \begin{equation}
      \gamma_n = \frac{n-2}{n-1}\,C_n\ .
   \end{equation}
   This parameter, as well as the characteristic length from Eq.~\eqref{eq:rEgen} has been introduced by \cite{STORMER_1907} in the particular context of $n=3$. As we will see, this allows us to describe all the possible trajectories in a unified way.\footnote{In the case $n=2$, the analogous parameter is $\gamma_2 = -\ln\rho_0$.}
   
\begin{figure*}
      \centering
      \includegraphics[width=0.7\textwidth]{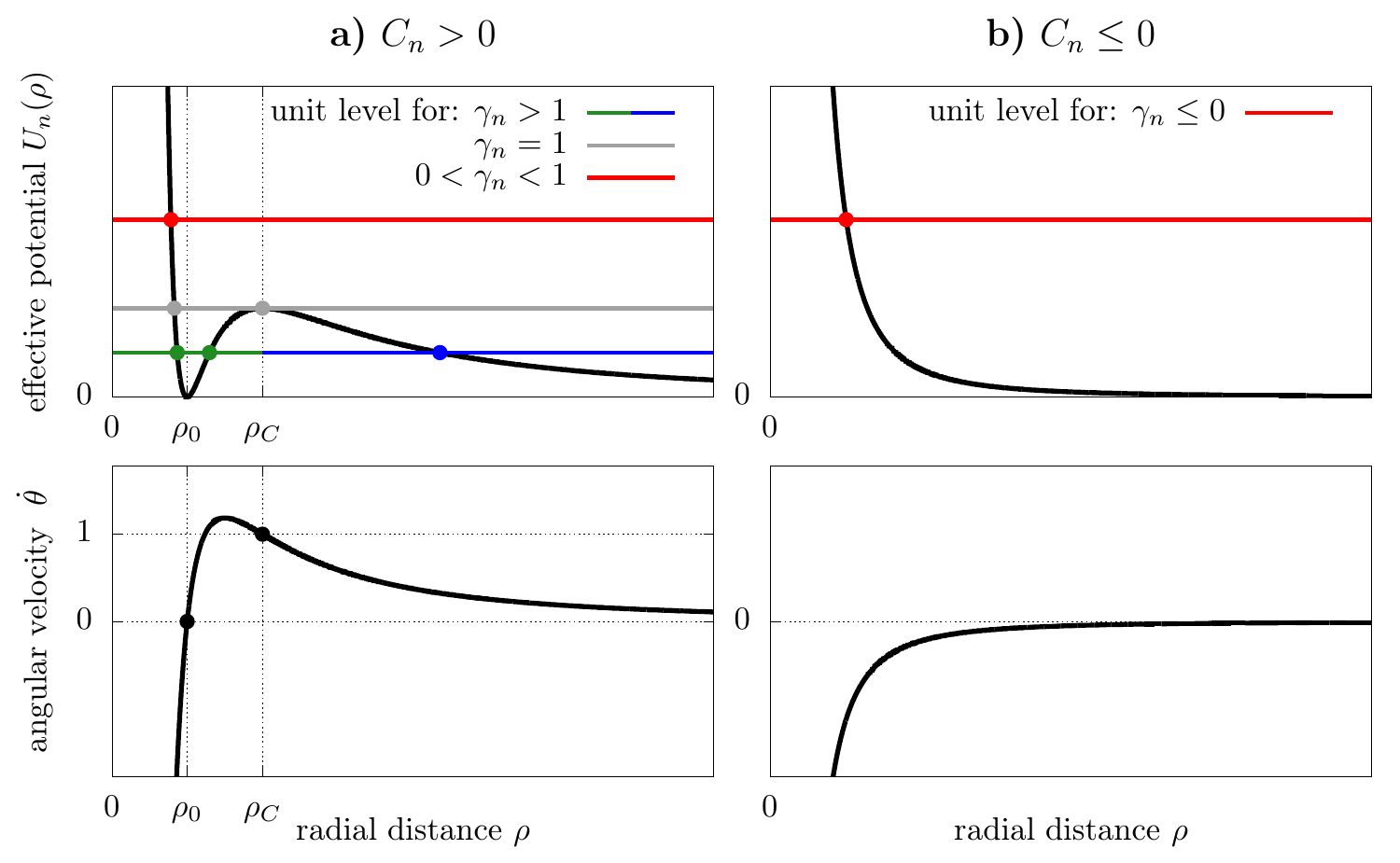}
      \caption{Effective potential and angular velocity as a function of $\rho$ in the three possible cases occurring for $n>2$. The unit level on the vertical axis gives the intervals of $\rho$ allowed for the particle, such that $U_n(\rho)<1$.}
      \label{fig:poteffn}
   \end{figure*}

   We must now consider the cases of negative, positive, and zero values of $C_n$. The effective potential and angular velocity as functions of $\rho$ are represented in Fig.~\ref{fig:poteffn} in the three cases. For $C_n>0$, the dynamics is pretty similar to the inverse-square-law field and we have the same types of trajectories. For $C_n\leqslant 0$, the only possible trajectories are of a type analogous to T$_1$, but for which the radius $\rho_0$ would be sent to infinity. Hence, their angular velocity is always negative.
   
   The extreme reachable radii are the positive roots of the two polynomials
   \begin{equation}\label{eq:roots}
      P_n^\pm(\rho) = \pm (n-2)\,\rho^{n-1} + (n-1)\,\gamma_n\,\rho^{n-2} - 1\ .
   \end{equation}
   From Descartes' rule of sign, we obtain that $P_n^+$ has exactly one positive root whatever the value of $\gamma_n$ (which defines $\rho_1$). On the other hand, $P_n^-$ has zero or two positive roots, and exactly zero if $\gamma_n<0$. For $\gamma_n>0$, $P_n^-$ has only one local maximum for $\rho>0$, equal to $\gamma_n^{n-1}-1$. Given that $P_n^-(0)=-1$ and $P_n^-(\infty)=-\infty$, we deduce as expected that $P_n^-$ has zero positive root if $\gamma_n<1$ and two if $\gamma_n\geqslant 1$ (which define $\rho_2$ and $\rho_3$). Since the polynomials are of order $n-1$, these roots have necessarily an explicit expression for $n\leqslant 5$ (Abel's impossibility theorem). For $n>5$, we have no guarantee that an explicit expression of the three radii exists, but they are still well-defined from Eq.~\eqref{eq:roots} and they can be determined numerically.
   
\begin{figure*}
      \centering
      \includegraphics[width=0.8\textwidth]{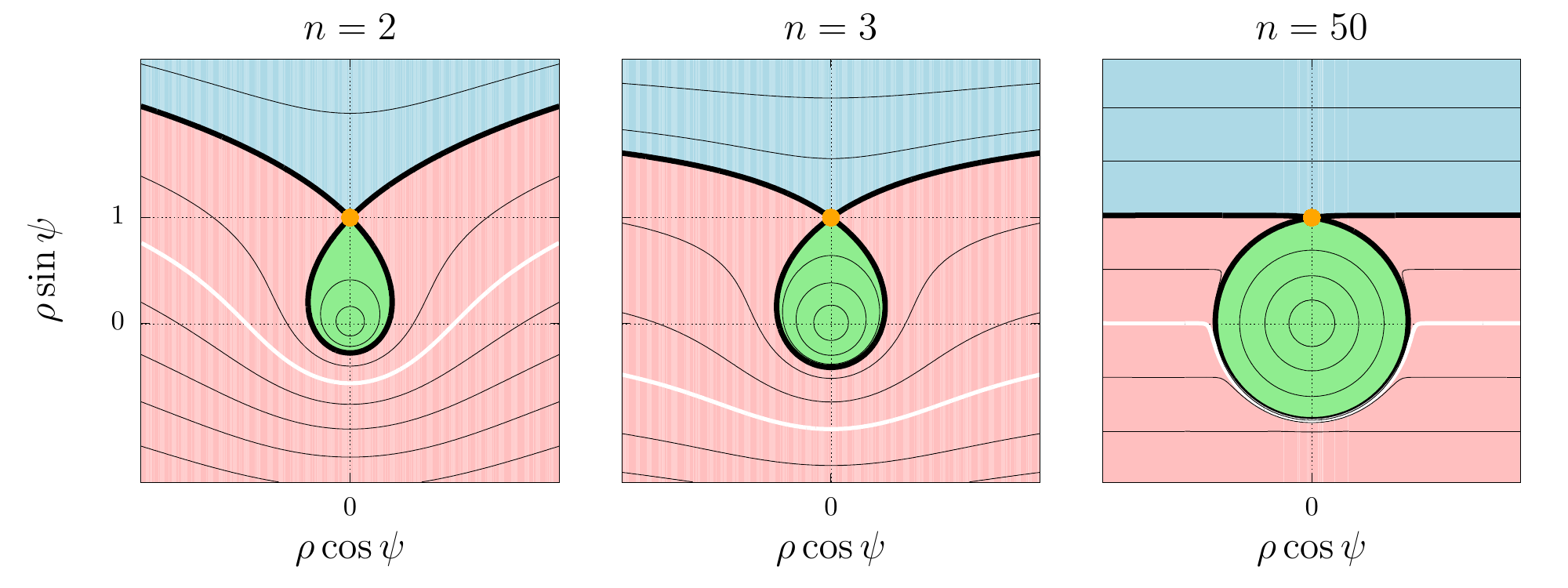}
      \caption{Phase portraits of the system in the plane $(\rho,\psi)$ for different powers $n$, obtained in terms of the level curves of $\gamma_n$ (with, in particular, $\gamma_2=-\ln\rho_0$). Trajectories of type T$_1$, T$_2$, and T$_3$ are plotted respectively in red, green, and blue. The thick black curve represents the unit level (separatrix). Along it lie the homoclinic orbit T$_2^\star$ and the two branches of the T$_3^\star$ orbit, whereas the circular trajectory T$^\star$ is plotted in orange. The white level curve represents the zero level. For $n>2$, it always remains in the $\rho\sin\psi < 0$ side.}
      \label{fig:phase}
   \end{figure*}

   Whatever the value of $n$, the different types of trajectories can be easily distinguished by plotting a phase portrait of the system. In our case, the best option is to use the level curves $\gamma_n$ in the plane $(\rho,\psi)$, where $\psi$ is the angle between the position and velocity vectors. Indeed, both $\dot{\rho}$ and $\dot{\theta}$ can be expressed in terms of $\psi$, leading to the following expressions:
   \begin{equation}
      \left\{
      \begin{aligned}
         \gamma_2 &= \rho\sin\psi - \ln\rho \\
         \gamma_n &= \frac{n-2}{n-1}\rho\sin\psi + \frac{1}{(n-1)\rho^{n-2}} \hspace{0.5cm},\hspace{0.5cm} n>2\ .
      \end{aligned}
      \right.
   \end{equation}
   The corresponding phase portraits are shown in Fig.~\ref{fig:phase}, showing that the dynamics in the cases $n\geqslant 2$ are qualitatively similar.
   
   If we consider an incoming flux of particles as in Sect.~\ref{sec:flux}, the $\gamma_n$ constant of the particles for $n>2$ is
   \begin{equation}\label{eq:gamma}
      \gamma_n(Y_i) = -\mathrm{sgn}(k)\,\frac{n-2}{n-1}\,Y_i + \frac{1}{n-1}\left(\frac{1}{\sqrt{D^2+Y_i^2}}\right)^{n-2} \,,
   \end{equation}
   and it is enough to study the case $k_n>0$. We note that
   \begin{equation}
      \lim\limits_{Y_i\rightarrow -\infty} \gamma_n(Y_i) = \infty
      \hspace{0.3cm}\text{and}\hspace{0.3cm}
      \lim\limits_{Y_i\rightarrow +\infty} \gamma_n(Y_i) = -\infty \,,
   \end{equation}
   so all the possible values of $\gamma_n$ are spanned by the initial positions $Y_i$, including the critical one $\gamma_n=1$. The study of $\gamma_n$ as a function of $Y_i$ and $D$ shows that the behaviour of the trajectories is qualitatively similar to what we obtained for $n=2$ (Sect.~\ref{sec:flux}). First of all, there is a limiting distance $D_\mathrm{lim}$ above which $\gamma_n$ is monotonous with respect to $Y_i$. It can be written in a very general way as
   \begin{equation}
      D_\mathrm{lim}^{(n)} = \sqrt{\frac{n-1}{n^\frac{n}{n-1}}}
      \hspace{0.5cm},\hspace{0.5cm}n>1\ .
   \end{equation}
   This formula is also valid for $n=2$. Then, there is a critical distance $D_\mathrm{crit}<D_\mathrm{lim}$ below which bounded trajectories appear (as in Fig.~\ref{fig:zones}). Finally the flux of incoming particles naturally creates a circular cavity similar to the case $n=2$. For $D>D_\mathrm{crit}$, the radius $\rho_\mathrm{cav}$ of this cavity is also independent of $D$: it is equal to the $\rho_1$ radius (Eq.~\ref{eq:roots}) at $\gamma_n=1$. We give in Tables~\ref{tab:Dcrit} and~\ref{tab:rhocav} the values of $D_\mathrm{crit}$ and $\rho_\mathrm{cav}$ for the first few $n$. We give also their analytical expression when we found one.
   
   \begin{table}[ht]
      \centering
      \begin{equation*}
         \begin{array}{c|c|c}
            n & \text{analytical }D_\mathrm{crit}^{(n)} & \text{numerical }D_\mathrm{crit}^{(n)} \\
            \hline
            2 & \frac{1}{2}\sqrt{-H\Big(H+2\Big)} & 0.4023711712747059 \\
            3 & \sqrt{2\big(I-2\big)} & 0.5193929104950238 \\
            4 & \frac{1}{2}\sqrt{6\sqrt{3}-9} & 0.5899798397854929 \\
            5 & unknown & 0.6389216906984257 \\
            6 & unknown & 0.6754661600886492 \\
            7 & unknown & 0.7040804466901285
         \end{array}
      \end{equation*}
      \caption{Critical starting distance below which bounded trajectories appear, given for the first few powers of $1/r$. We define $H = W_0(-2\exp[-2])$ and $I = \sqrt[3]{\frac{11\sqrt{33}}{9}+7} - \sqrt[3]{\frac{11\sqrt{33}}{9}-7}$. The expression for $n=2$ is taken from Sect.~\ref{sec:flux}.}
      \label{tab:Dcrit}
   \end{table}
   
   \begin{table}[ht]
      \centering
      \begin{equation*}
         \begin{array}{c|c|c}
            n & \text{analytical }\rho_\mathrm{cav}^{(n)} & \text{numerical }\rho_\mathrm{cav}^{(n)} \\
            \hline
            2 & W_0(\exp[-1])     & 0.278464542761074 \\
            3 & \sqrt{2} - 1 & 0.414213562373095\\
            4 & 1/2          & 0.5 \\
            5 & -\frac{1}{3} - J + \sqrt{\frac{1}{3} - J^2 + \frac{2}{27}\frac{1}{J}} & 0.560425660450317 \\
            6 & \frac{1}{4}\big(1+K\sqrt[3]{5}\big) & 0.605829586188268 \\
            7 & unknown & 0.641465469828847
         \end{array}
      \end{equation*}
      \caption{Radius of the cavity formed by a flux of particles coming from a distance $D>D_\mathrm{crit}^{(n)}$, for the first few powers of $1/r$. We define $J = \frac{1}{6}\sqrt{4 + 6\sqrt[3]{\sqrt{2}-1} - 6\sqrt[3]{\sqrt{2}+1}}$ and $K = \sqrt[3]{\frac{4\sqrt{6}}{9}+1}-\sqrt[3]{\frac{4\sqrt{6}}{9}-1}$. The expression for $n=2$ is taken from Sect.~\ref{sec:flux}.}
      \label{tab:rhocav}
   \end{table}
   
\begin{figure*}
      \centering
      \includegraphics[width=\textwidth]{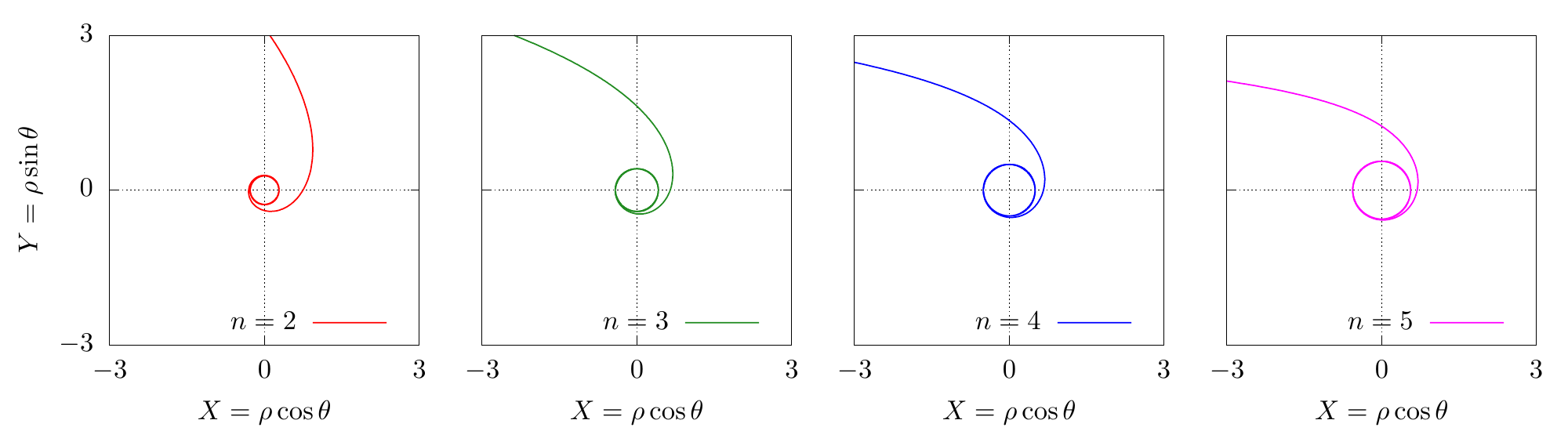}
      \caption{Comparison of the caustics obtained for the first few values of $n$. We recognise the sizes of the central cavities given in Table~\ref{tab:rhocav}.}
      \label{fig:causticN}
   \end{figure*}

   Eventually, one can use the implicit solution obtained by quadrature in order to compute the shape of the caustic, as we did in Sect.~\ref{sec:flux}. For $n>2$, we get
   \begin{equation}
      \left\{
      \begin{aligned}
         \varphi_n(\rho) &= \frac{(n-1)\gamma_n\,\rho^{n-2}-1}{\rho\sqrt{\big((n-2)\rho^{n-1}\big)^2-\big((n-1)\gamma_n\,\rho^{n-2}-1\big)^2}} \\
         \phi_n(\rho) &= \frac{(n-2)\rho^{n-1}}{\sqrt{\big((n-2)\rho^{n-1}\big)^2-\big((n-1)\gamma_n\,\rho^{n-2}-1\big)^2}} \,.
      \end{aligned}
      \right.
   \end{equation}
   These functions can be used directly as in Eqs.~\eqref{eq:traj} and \eqref{eq:tauR}, with the same parametrisation $s\in\mathbb{R}$. If we consider particles coming from infinity $(D\rightarrow\infty)$, there is a notable difference with respect to the case $n>2$. Indeed, the parameter $\gamma_n$ of the particles (Eq.~\ref{eq:gamma}) becomes directly proportional to $Y_i$. It is thus much simpler than in Sect.~\ref{sec:flux}, since $Y_i$ now keeps a clear meaning even when $D$ is infinite: it becomes the impact parameter of the particles. For completeness, Fig.~\ref{fig:causticN} compares the shape of the caustics obtained for the first few values of $n$.
   
\end{document}